\documentclass[11pt,a4]{article}
\usepackage{amsfonts}
\usepackage{graphics,amssymb,amsmath,epsf}
\usepackage{graphicx}
\usepackage{subfigure}

\numberwithin{equation}{section}

\begin{document}

\date{}
\title{Beta-Linear Failure Rate Distribution and its Applications}
\author{A. A. Jafari, E. Mahmoudi\thanks{corresponding; Email:emahmoudi@yazd.ac.ir; Phone: +98-351-8122703;
 Fax: +98-351-8210695} \\
Department of Statistics, Yazd University, Yazd, Iran}
\maketitle

\begin{abstract}
We introduce in this paper a new four-parameter generalized version
of the linear failure rate (LFR) distribution which is called
Beta-linear failure rate (BLFR) distribution. The new distribution
is quite flexible and can be used effectively in modeling survival
data and reliability problems. It can have a constant, decreasing,
increasing, upside-down bathtub (unimodal) and bathtub-shaped
failure rate function depending on its parameters. It includes some
well-known lifetime distributions as special sub-models. We provide
a comprehensive account of the mathematical properties of the new
distributions. In particular, A closed-form expressions for the
density, cumulative distribution and hazard rate function of the
BLFR is given. Also, the $r$th order moment of this distribution is
derived. We discuss maximum likelihood estimation of the unknown
parameters of the new model for complete sample and obtain an
expression for Fisher�s information matrix. In the end, to show the
flexibility of this distribution and illustrative purposes, an
application using a real data set is presented.

\bigskip



\end{abstract}

\section{Introduction}

The linear failure rate distribution with parameters $a\geq0$ and
$b\geq0$, ($a+b>0$) which is denoted by ${\rm LFRD}(a,b)$, has the cumulative
distribution function (CDF)
\begin{equation}\label{eq.FLFR}
G\left(x\right)=1-{\exp  \left(-ax-\frac{b}{2}x^2\right)},\ \ \ \ \ x>0,
\end{equation}
and probability density function
\begin{equation}\label{eq.fLFR}
{\rm g}\left(x\right)=\left(a+bx\right){\exp  \left(-ax-\frac{b}{2}x^2\right)},\ \ \ \ \ x>0.
\end{equation}

Note that if $b=0$ and $a\neq0$, then the LFR distribution is
reduced to exponential distribution with parameter $a$ (${\rm
Exp}(a)$), and if $a=0$ and $b\neq0$ then we can obtain the Rayleigh
distribution with parameter $b$ (${\rm Rayleigh}(b)$). A basic
structural properties of ${\rm LFRD(}a,b)$ is that it is the
distribution of minimum of two independent random variables $X_1$
and $X_2$ having ${\rm Exp}(a)$ and ${\rm Rayleigh}(b)$
distributions, respectively (Sen and Bhattachrayya, 1995).

If $G$ denotes the CDF  of a random variable then a generalized class of distributions can be defined by
\begin{equation} \label{eq.FB}
F\left(x\right)=I_{G\left(x\right)}\left(\alpha ,\beta \right)=\frac{1}{B\left(\alpha ,\beta \right)}\int^{G\left(x\right)}_0{t^{\alpha -1}{\left(1-t\right)}^{\beta -1}}dt,
\end{equation}
for $\alpha >0$ and $\beta >0$, where $I_y\left(\alpha ,\beta \right)=\frac{B_y\left(\alpha ,\beta \right)}{B\left(\alpha ,\beta \right)}$ is the incomplete beta function ratio and $B_y\left(\alpha ,\beta \right)=\int^y_0{t^{\alpha -1}{\left(1-t\right)}^{\beta -1}}$ is the incomplete beta function.

Many authors considered various forms of $G$ and studied their properties: Eugene et al. (2002) (Beta Normal distribution), Nadarajah and Kotz (2004) (Beta Gumbel distribution), Nadarajah and Gupta (2004) and Barreto-Souza et al. (2011)  (Beta Fr$\acute{{\rm e}}$chet distribution), Famoye et al. (2005), Lee et al. (2007)  and Cordeiro et al. (2008) (Beta Weibull distribution), Nadarajah and Kotz (2006) (Beta Exponential distribution), Akinsete et al. (2008) (Beta Pareto distribution), Silva et al. (2010) (Beta Modified Weibull distribution), Mahmoudi (2011) (Beta generalized Pareto distribution), Cordeiro et al. (2011) (Beta-exponentiated Weibull distribution), Cordeiro et al. (2011) (Beta-Weibull geometric distribution), Singla et al. (2012) (Beta generalized Weibull distribution), Cordeiro et al. (2012) (Beta generalized gamma distribution) and Cordeiro et al. (2012) (Beta generalized normal distribution).

In this article, we propose a new four parameters distribution,
referred to as the BLFR distribution, which contains as special
sub-models: the Beta exponential (BE), Beta Rayleigh (BR),
generalized linear failure rate (GLFR) and linear failure rate (LFR)
distributions, among others. The main reasons for introducing BLFR
distribution are: (i) The additional parameters introduced by the
beta generalization is sought as a means to furnish a more flexible
distribution. (ii) Some modeling phenomenon with non-monotone
failure rates such as the bathtub-shaped and unimodal failure rates,
which are common in reliability and biological studies, take a
reasonable parametric fit with this distribution. (iii) The BLFR
distribution is expected to have immediate application in
reliability and survival studies. (iv) BLFR distribution shows
better fitting, more flexible in shape and easier to perform and
formula for modeling lifetime data.

The reminder of the paper is organized as follows: In Section 2, we define the BLFR distribution and outline some special cases of the distribution. We investigate some properties of the distribution in this Section. Some of these properties are the limit behavior and shapes of the pdf and hazard rate function of the BLFR distribution. Section 3 provides a general expansion for the moments of the BLFR distribution. In Section 4, we discuss maximum likelihood estimation and calculate the elements of the observed information matrix. Application of the BLFR distribution is given in the Section 5. A simulation study is performed in Section 6. Finally, Section 7 concludes the paper.

\section{ Definition of the BLFR distribution and some special cases}

Consider that ${\rm g}(x)\ =\ dG(x)/dx$ is the density of the
baseline distribution. Then the probability density function
corresponding to (\ref{eq.FB}) can be written in the form
\begin{equation} \label{eq.fB}
f\left(x\right)=\frac{g(x)}{B\left(\alpha ,\beta
\right)}{G\left(x\right)}^{\alpha
-1}{\left(1-G\left(x\right)\right)}^{\beta-1}.
\end{equation}

We now introduce the BLFR distribution by taking $G(x)$ in (\ref{eq.FB}) to be the CDF (\ref{eq.FLFR}) of the LFR distribution. Hence, the BLFR density function can be written as
\begin{equation} \label{eq.fBLFR}
f\left(x\right)=\frac{a+bx}{B\left(\alpha ,\beta
\right)}{\left(1-{\exp  \left(-ax-\frac{b}{2}x^2\right)\
}\right)}^{\alpha-1}{\exp  \left(-a\beta x-\frac{b\beta}{2}x^2\right) },
\end{equation}
and we use the notation $X\sim {\rm BLFR}\ (a, b, \alpha , \beta )$.

The hazard rate function of BLFR distribution is given by
\begin{equation} \label{eq.hBLFR}
h\left(x\right)=\frac{a+bx}{B\left(\alpha ,\beta
\right)-B_{G(x)}\left(\alpha ,\beta \right)}{\left(1-{\exp
\left(-ax-\frac{b}{2}x^2\right)\ }\right)}^{\alpha -1}{\exp
\left(-a\beta x-\frac{b\beta }{2}x^2\right) }.
\end{equation}

\begin{figure}[]
\centering
\includegraphics[scale=0.45]{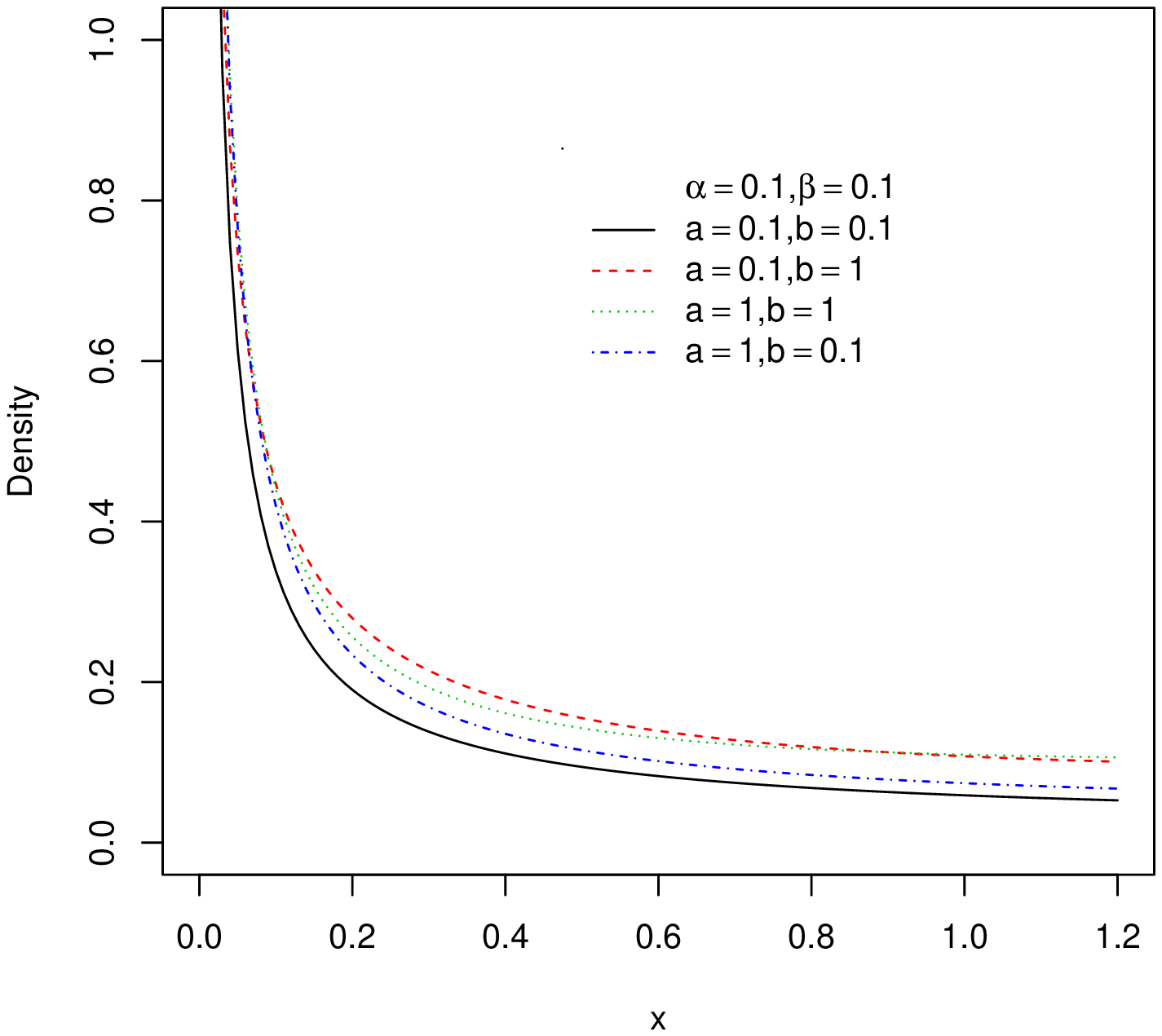}
\includegraphics[scale=0.45]{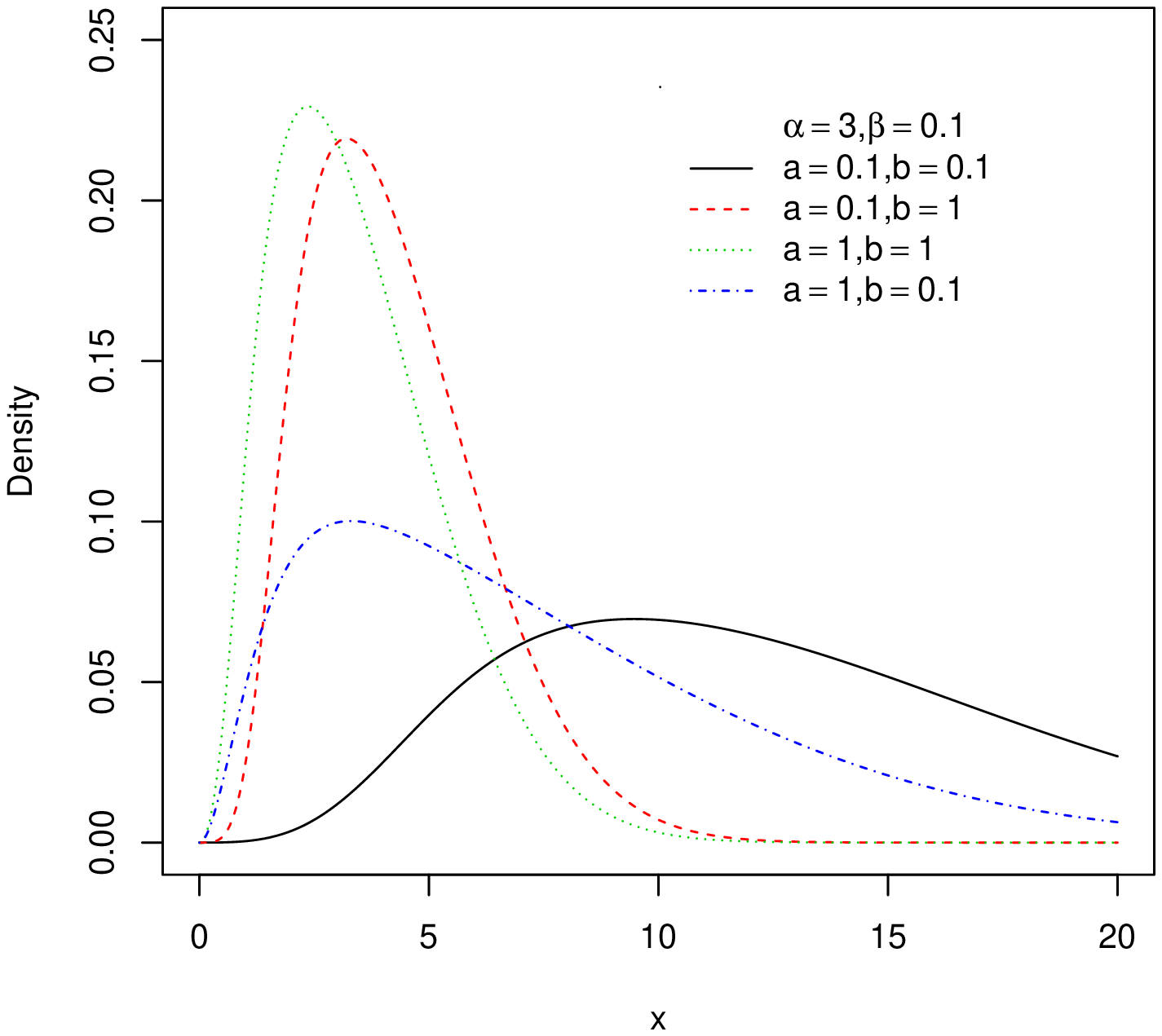}
\includegraphics[scale=0.45]{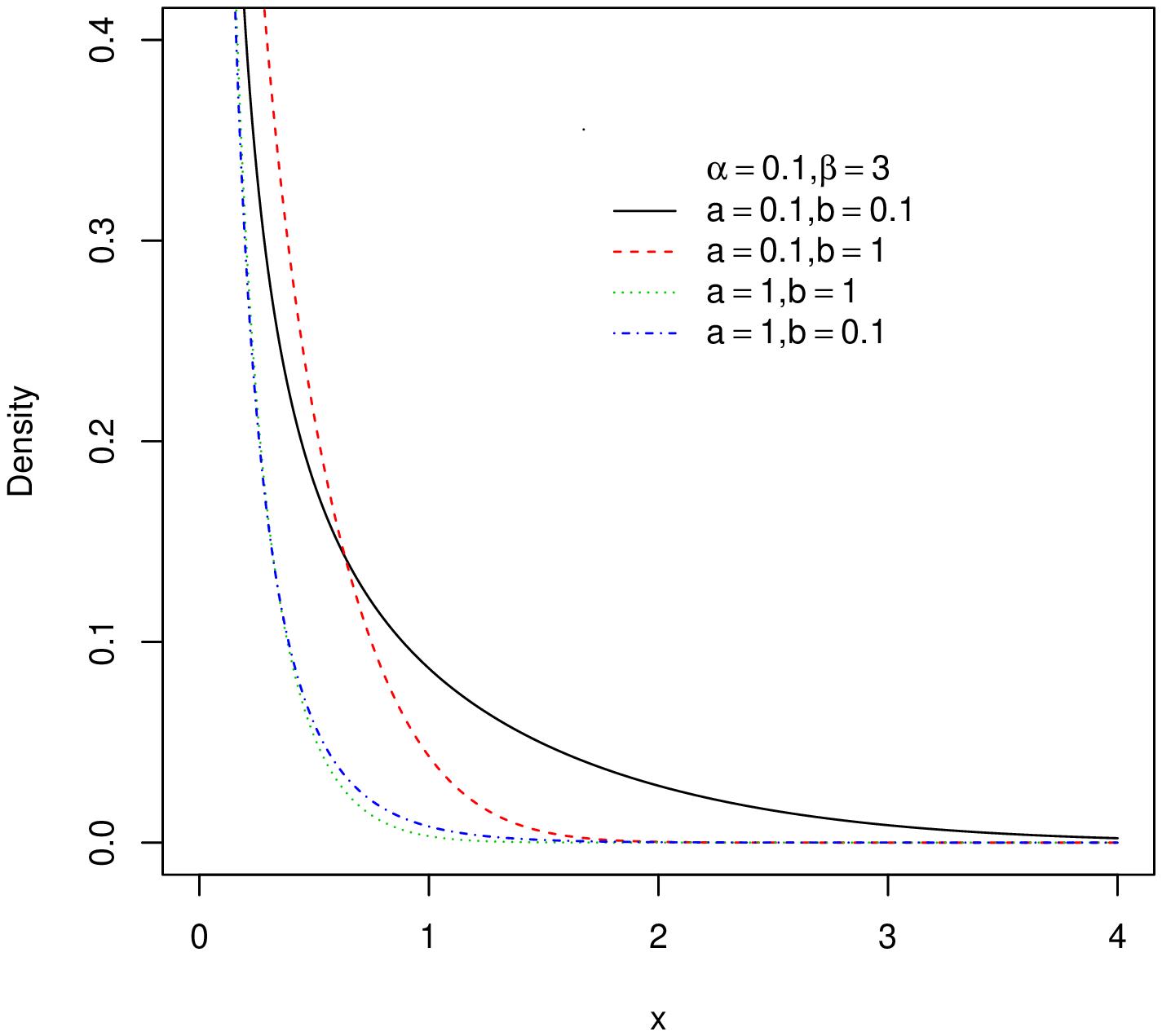}
\includegraphics[scale=0.45]{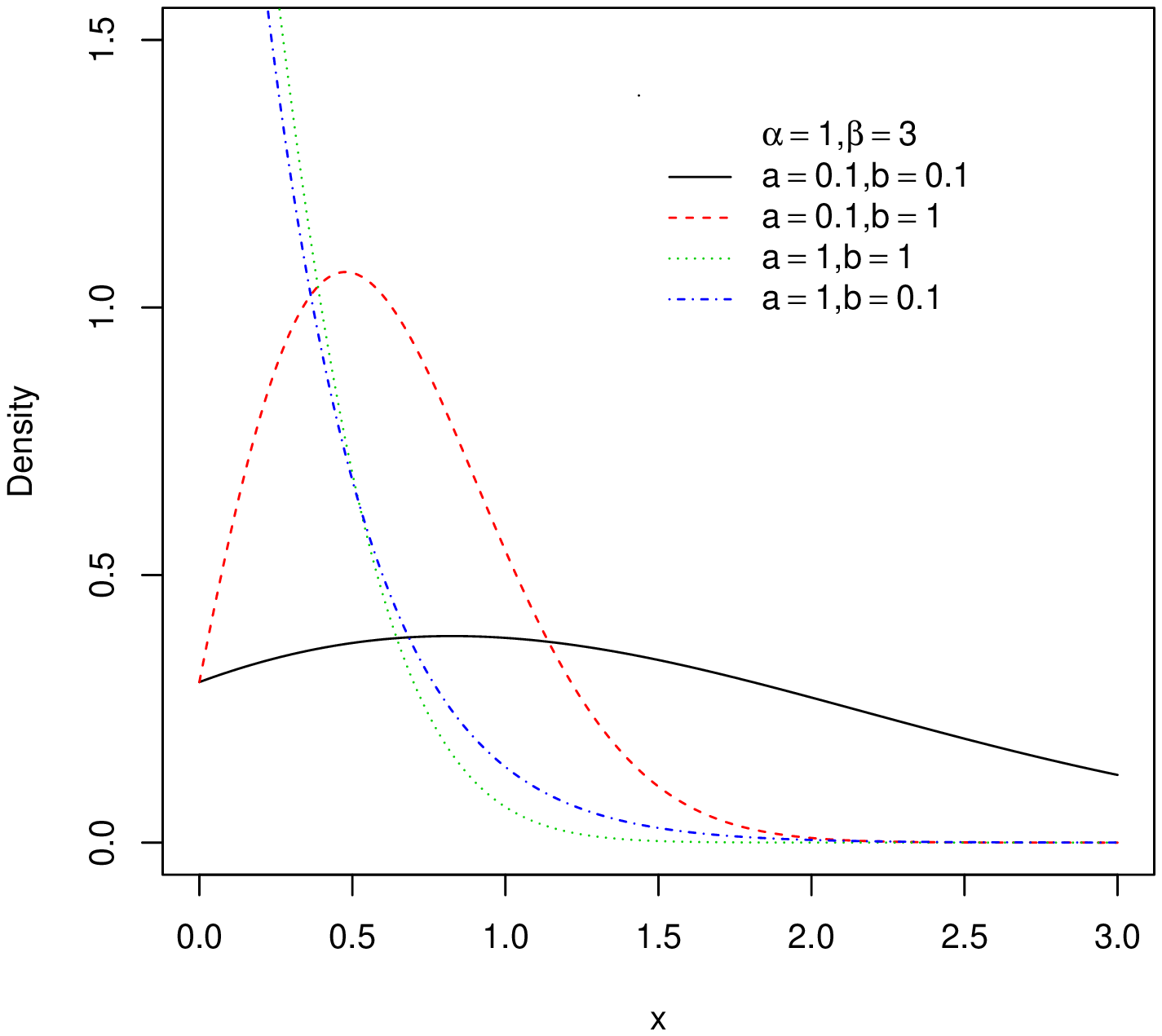}
\includegraphics[scale=0.40]{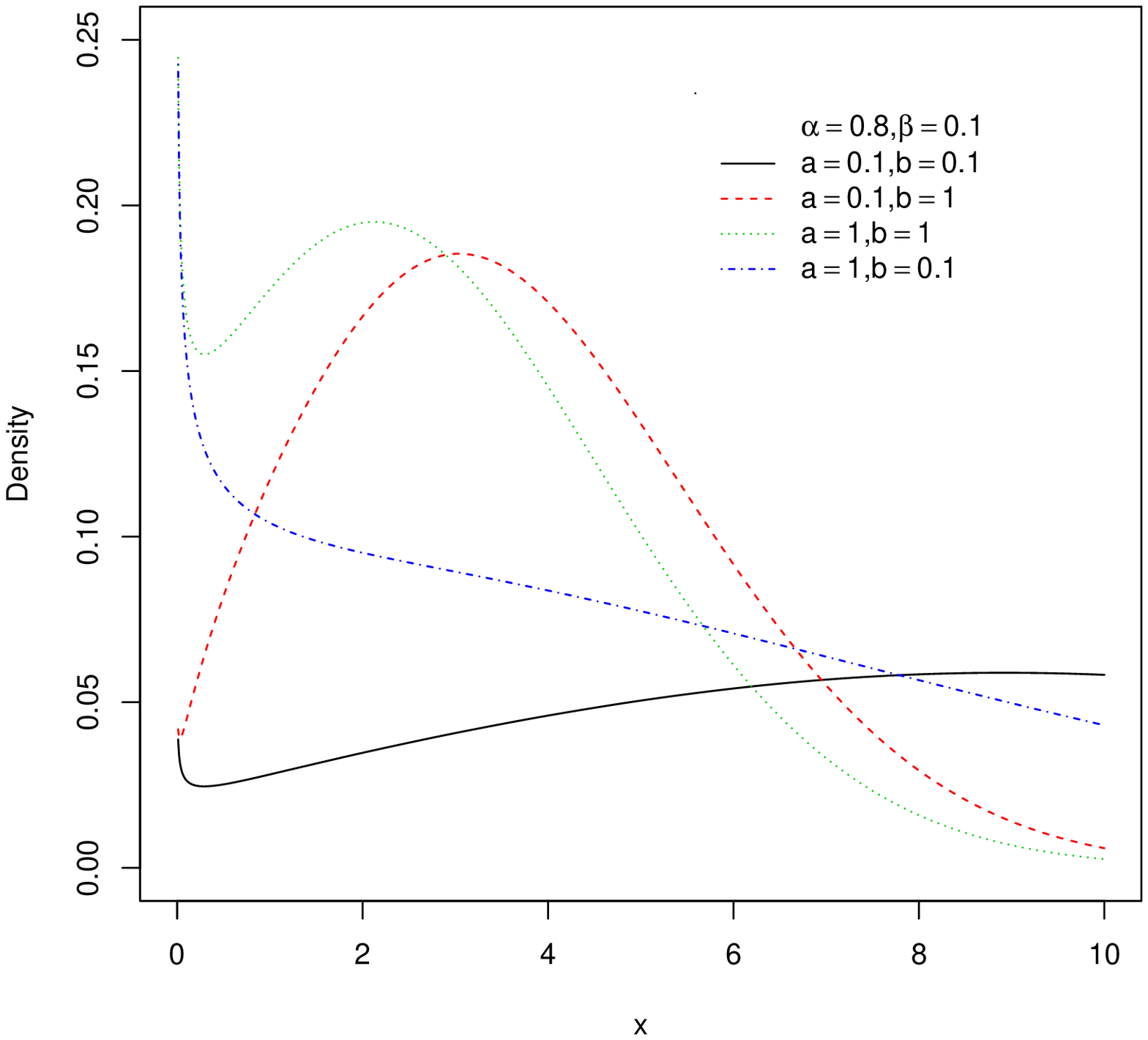}
\includegraphics[scale=0.40]{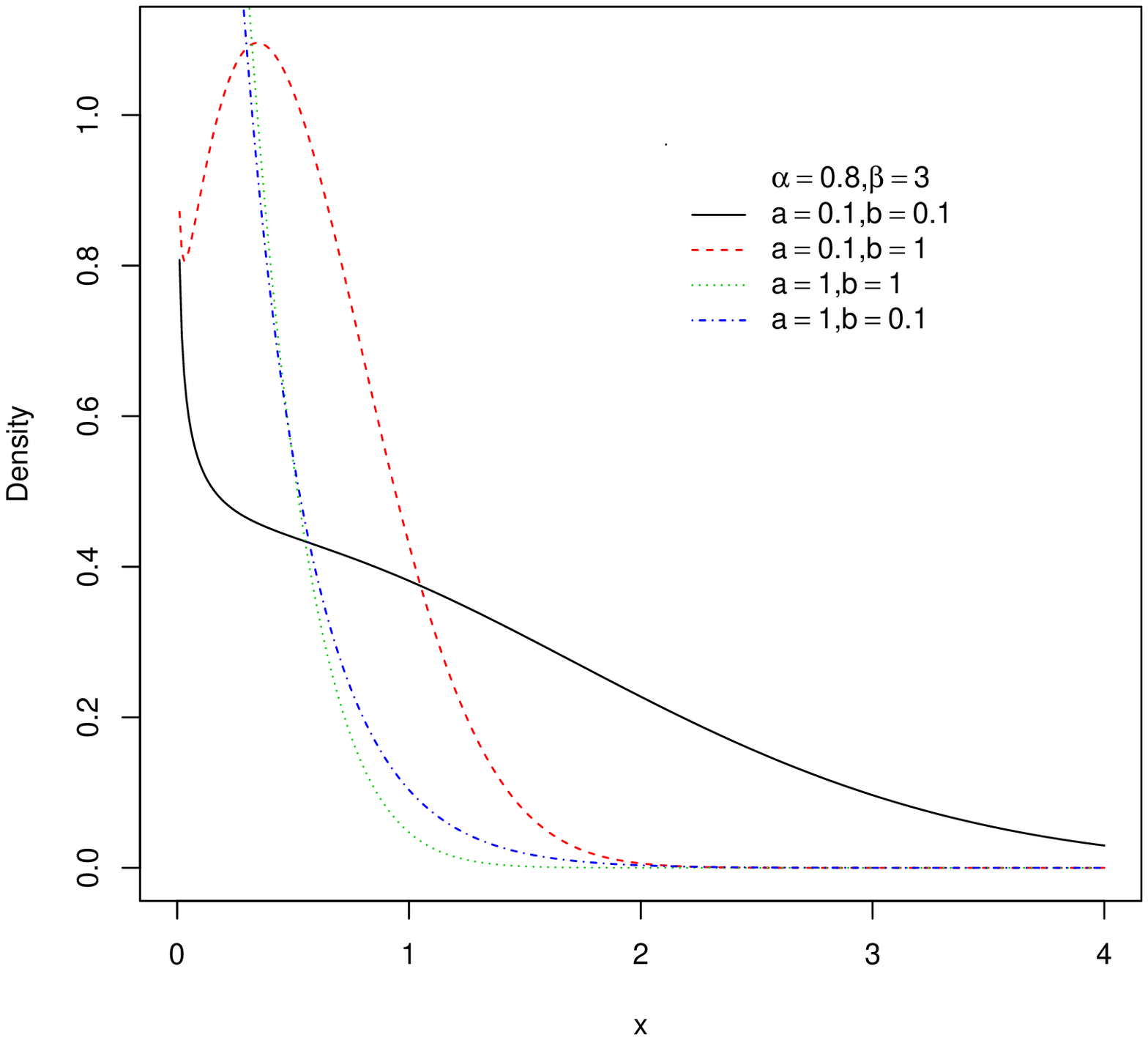}
\caption[]{Plots of pdf of the BLFR distribution for selected
parameters. }\label{fig.den}
\end{figure}

\begin{figure}[]
\centering
\includegraphics[scale=0.45]{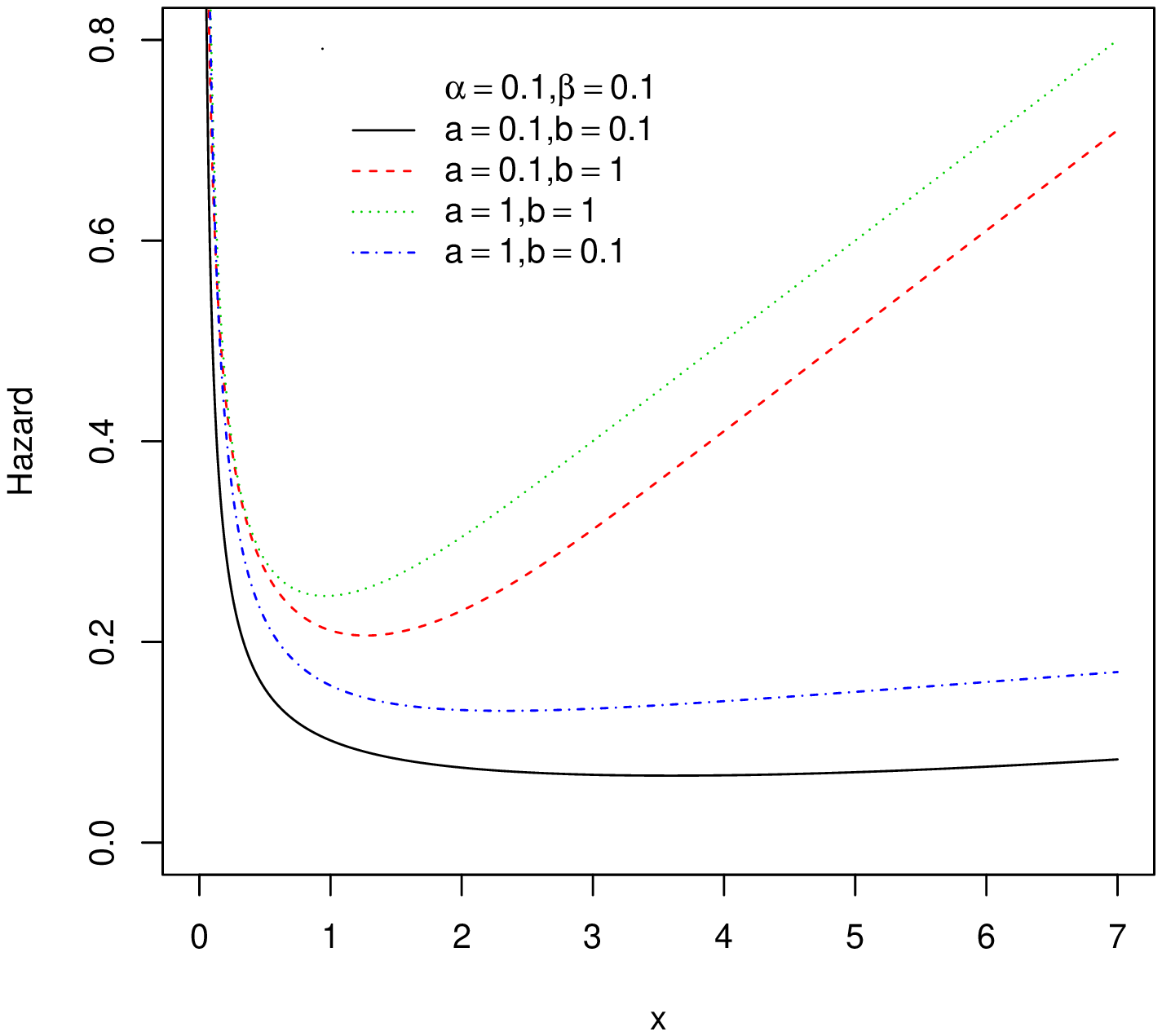}
\includegraphics[scale=0.45]{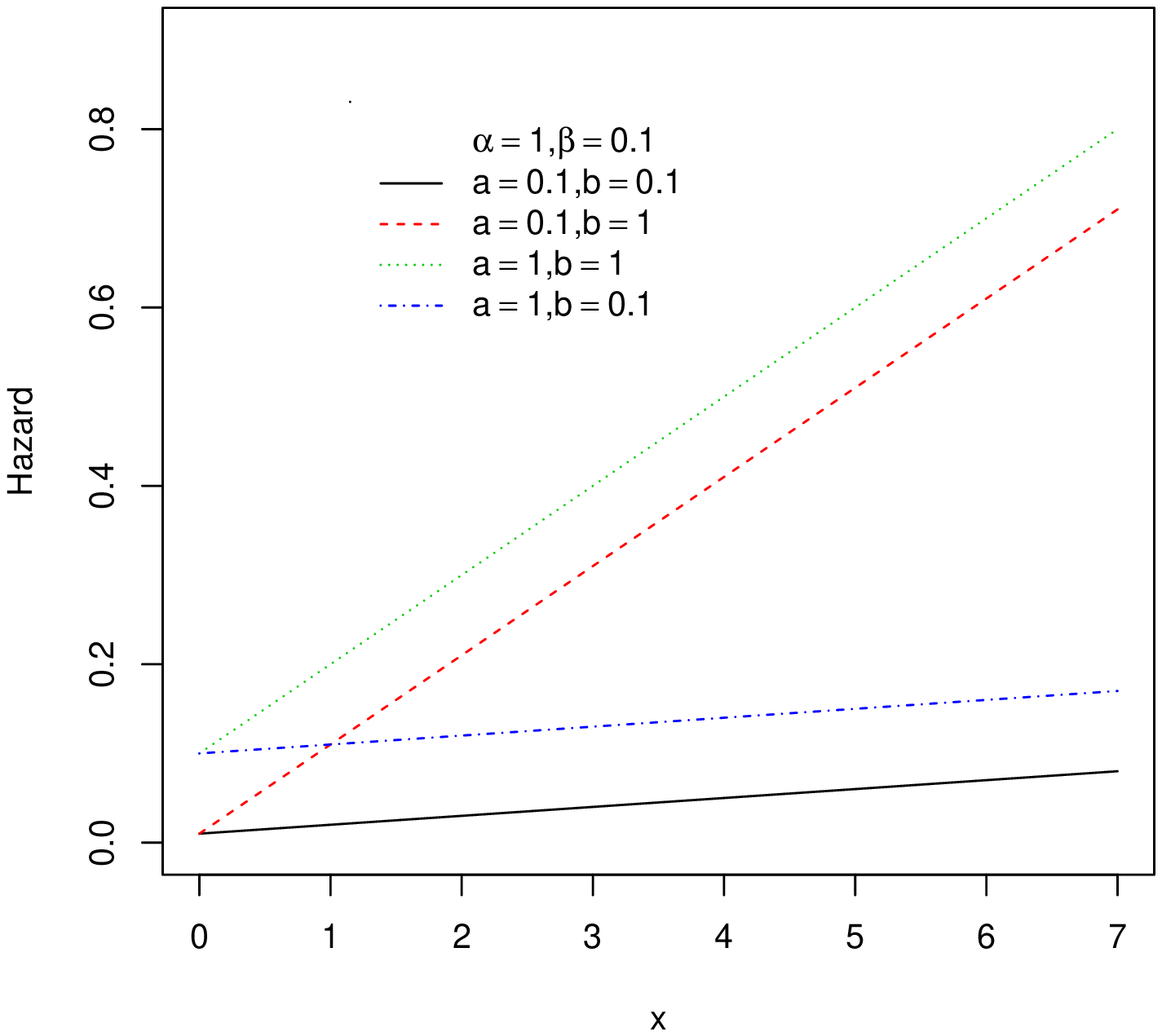}
\includegraphics[scale=0.45]{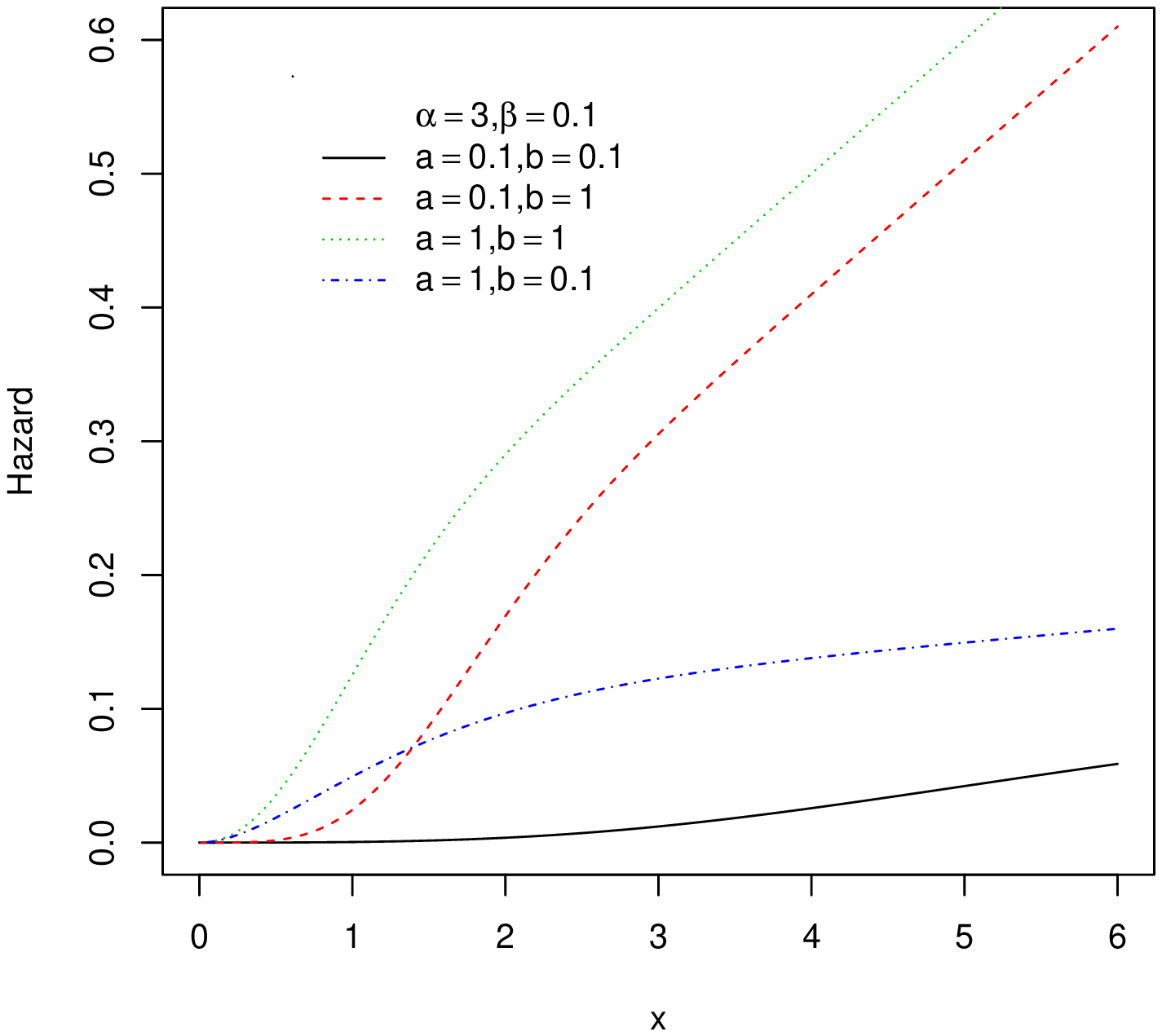}
\includegraphics[scale=0.45]{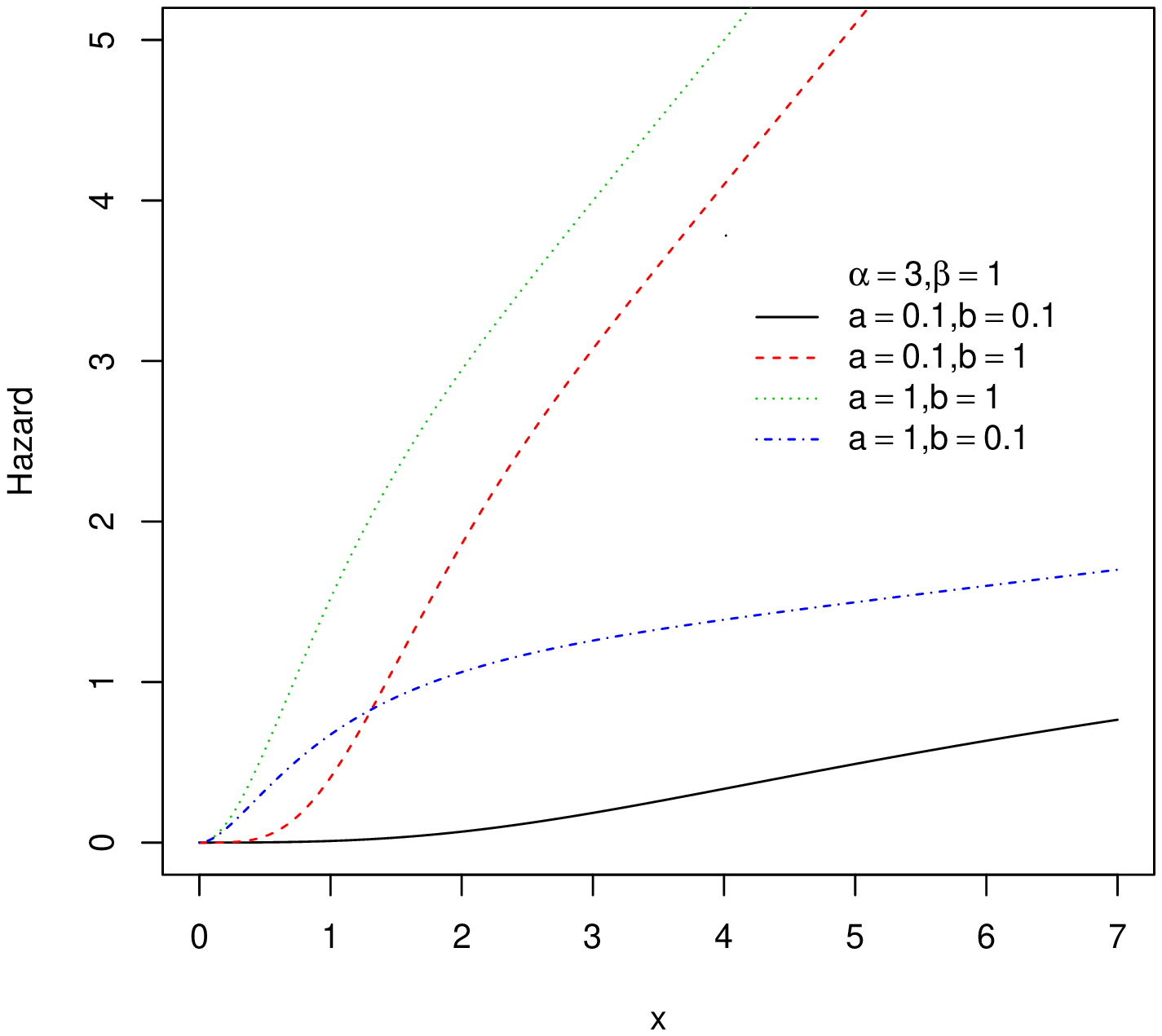}
\includegraphics[scale=0.45]{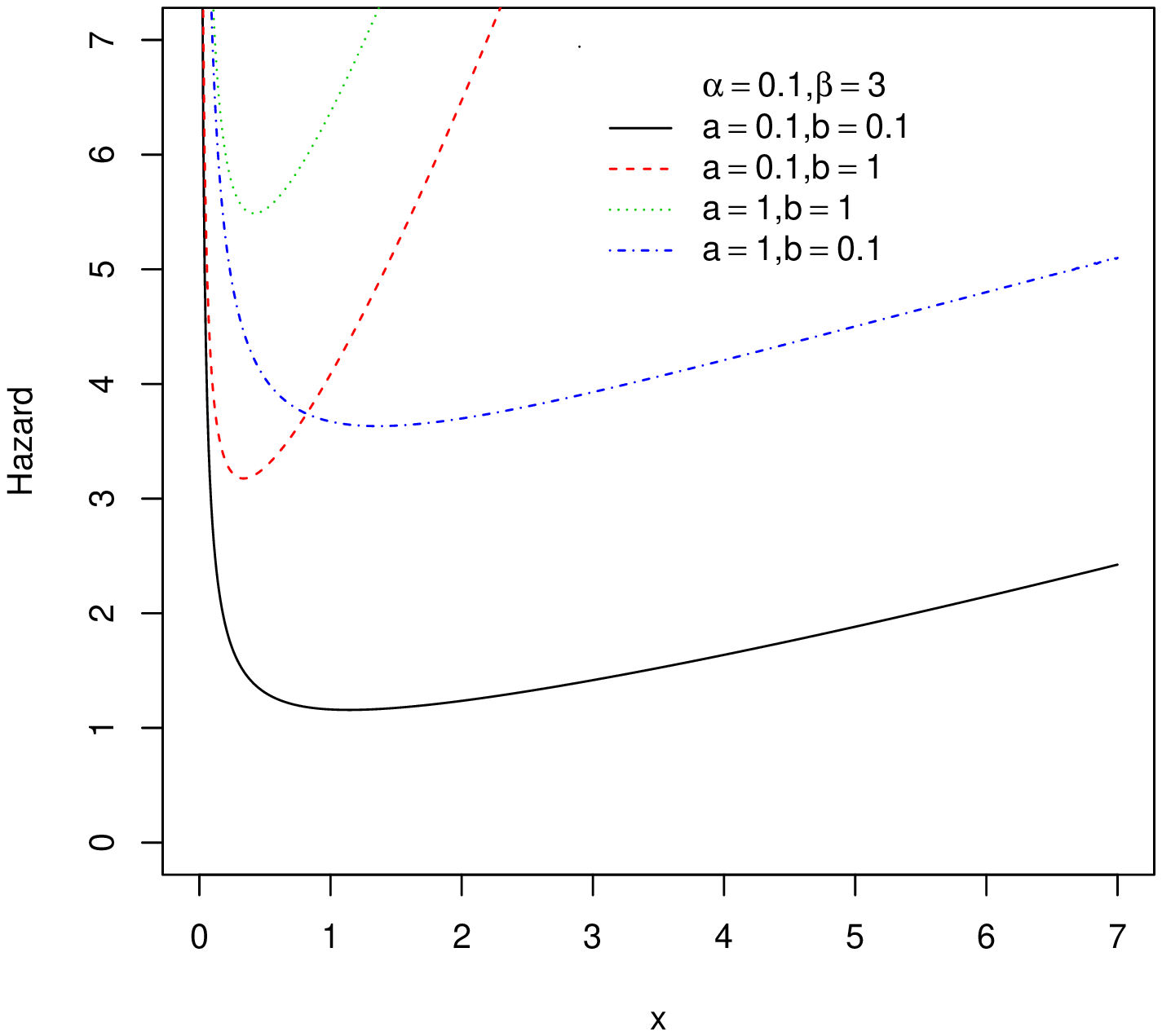}
\includegraphics[scale=0.45]{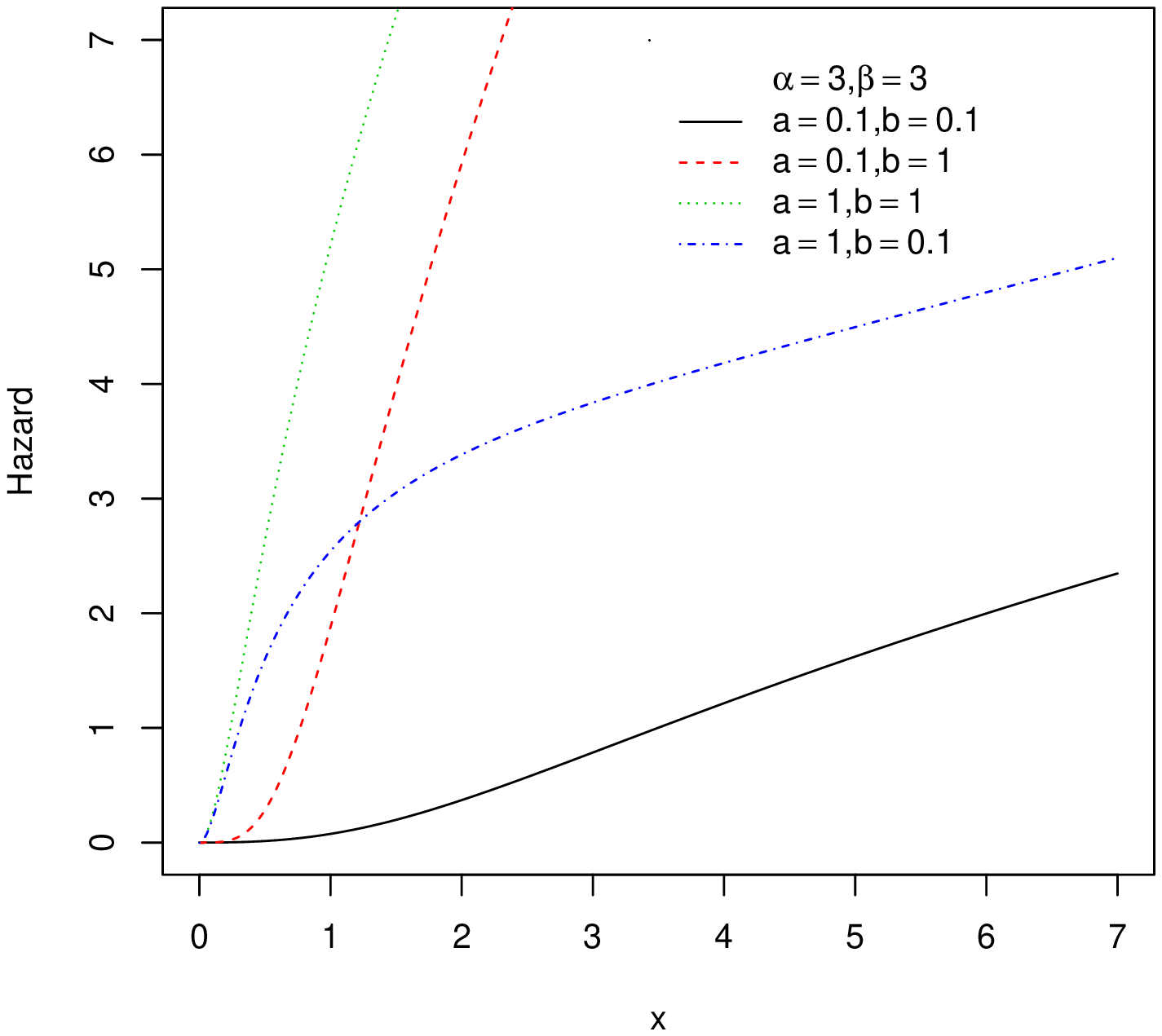}
\caption[]{Plots of hazard rate function of the BLFR distribution
for selected parameters. } \label{fig.hz}
\end{figure}

Plots of pdf and hazard rate function of the BLFR distribution for
different values of it's parameters are given in Fig. \ref{fig.den}
and Fig. \ref{fig.hz}, respectively.

\subsection{ Special cases of the BLFR distribution}

\begin{enumerate}
\item If $\beta =1$, then we get the generalized linear failure rate distribution (${\rm GLFR}(a,b,\alpha )$) which is introduced by Sarhan and Kundu (2009).

\item If $\beta =1$ and $b=0$, then we get the generalized exponential distribution (GE) (Gupta and Kundu, 1999).

\item If $\beta =1$ and $a=0$, then we get two-parameter Burr X distribution which is introduced by Surles and Padgett (2005) and also is known as generalized Rayleigh distribution (GR) (Kundu and Raqab, 2005) .

\item If $\alpha =\beta =1$, then (2.2) reduces to the linear failure rate distribution (${\rm LFR}(a,b)$) distribution.

\item If $b=0$, then we get the  beta exponential distribution (${\rm BE}(a,\alpha ,\beta )$) which is introduced by Nadarajah and Kotz (2006).

\item If $a=0$, then we get the  beta Rayleigh distribution (${\rm BR}(b,\beta )$) which is defined by Akinsete and Lowe (2009) and is a special case of beta Weibull distribution (Famoye et al., 2005).

\item If the random variable $X$\textit{ }has BLFR distribution, then the random variable
\[Y=1-{\exp  \left(-aX-\frac{b}{2}X^2\right)\ },\]
satisfies the beta distribution with parameters $\alpha $ and $\beta $. Therefore,
\[T=aX+\frac{b}{2}X^2\]
satisfies the beta exponential distribution with parameters 1, $\alpha $ and $\beta $ (${\rm BE}(1,\alpha ,\beta )$).

\item If $\alpha =i$ and $\beta =n-i$, where $i$ and $n$ are positive integer values, then the $f(x)$ is the density function of $i$th order statistic of LFR distribution.

\end{enumerate}

The following result helps in simulating data from the BLFR
distribution: If $Y$  follows Beta distribution with parameters
$\alpha $ and $\beta $, then
\[X=G^{-1}\left(Y\right)=\left\{ \begin{array}{lcc}
\frac{-a+\sqrt{a^2-2b{\log  (1-Y)\ }}}{b} &  & {\rm if}\ \ \  b>0 \\
 &  &  \\
-\frac{{\log  \left(1-Y\right)\ }}{a}\ \ \ \ \ \   &  & {\rm \ \ \  \ \ \ \ if\ \  \ }a{\rm >}0,\ b=0, \end{array}
\right.\]
follows BLFR distribution with parameters $a,\ b,\ \alpha $, and $\beta $.

\begin{figure}[ht]
\centering
\includegraphics[scale=0.45]{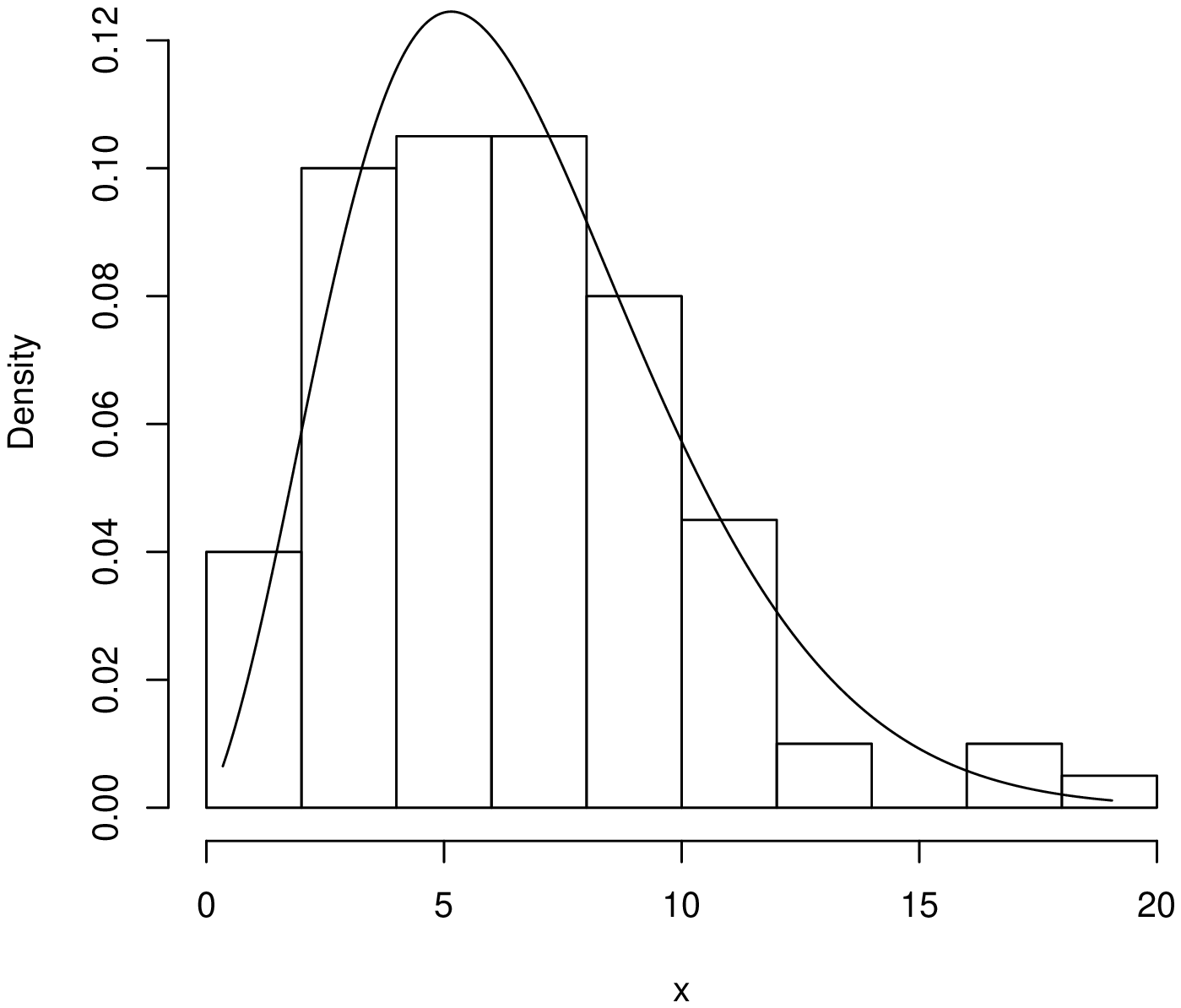}
\includegraphics[scale=0.45]{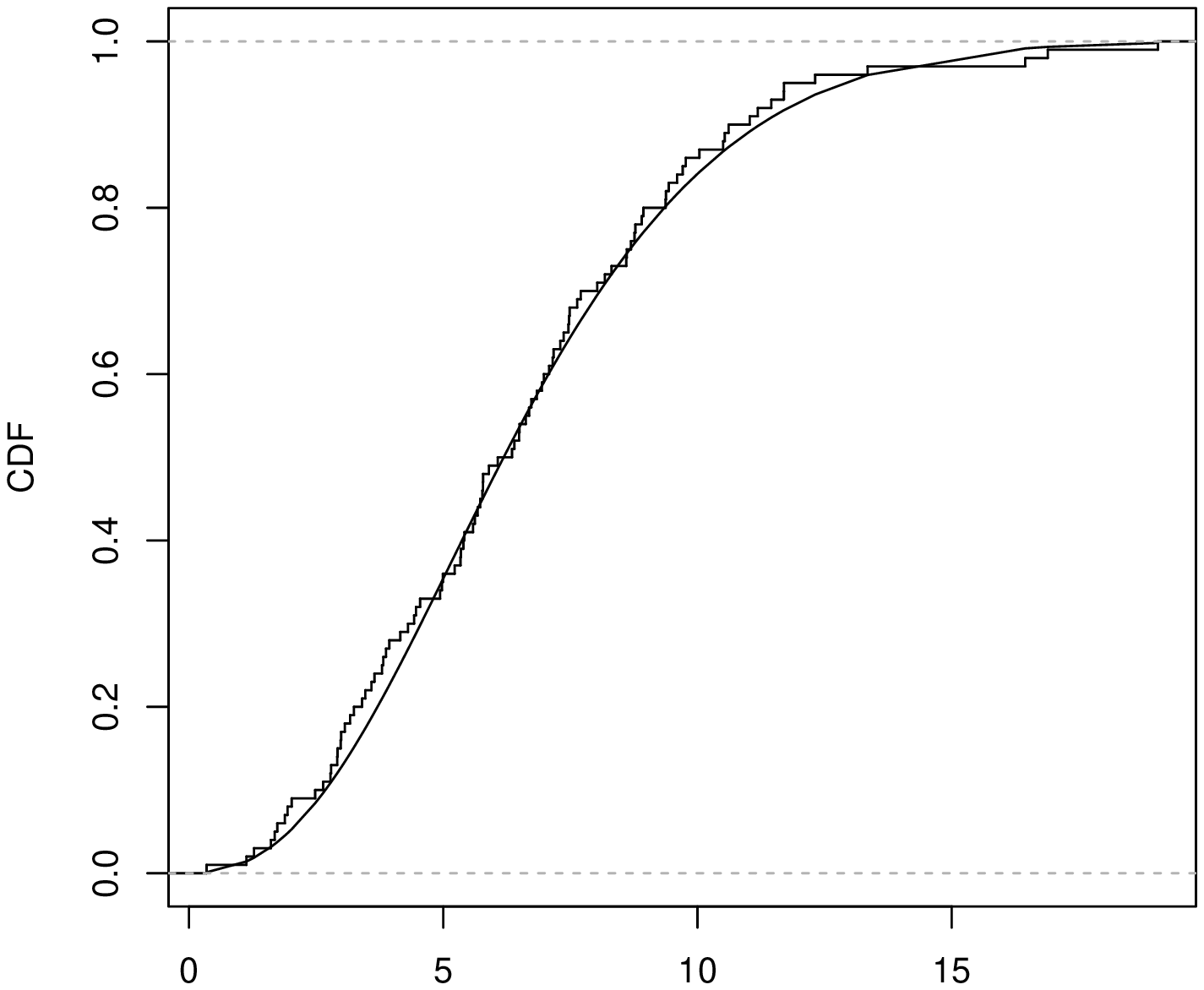}
\caption[]{The histogram of a generated data set with size 100 and
the exact BLFR density (left) and the empirical distribution
function and exact distribution function (right). } \label{Fig.gd}
\end{figure}

For checking the consistency of the simulating data set form BLFR
distribution, the histogram for a generated data set with size 100
and the exact BLFR density with parameters $a=0.2$, $b=0.1$,
$\alpha=2$, and $\beta=0.3$,  are displayed in Fig \ref{Fig.gd}
(left). Also, the empirical distribution function and the exact
distribution function is given in Fig \ref{Fig.gd} (right).

\subsection{ Properties of the BLFR distribution}

In this section, limiting behavior of pff and hazard rate function
of the BLFR distribution and their shapes are studied.

\bigskip

\noindent {\bf Theorem 1.} Let $f(x)$ be the pdf of the BLFR
distribution. The limiting behaviour of $f$ for different values of
its parameters is given bellow:
\begin{description}
\item[i.] If $\alpha =1$ then ${\mathop{\lim }_{x\rightarrow 0} f(x)\ }=a\beta$.

\item[ii.] If $\alpha >1$ then ${\mathop{\lim }_{x \rightarrow 0} f(x)\ }=0$.

\item[iii.] If $0<\alpha <1$ then ${\mathop{\lim }_{x \rightarrow 0} f(x)\ }=\infty$.

\item[iv.] ${\mathop{\lim }_{x\rightarrow \infty} f(x)\ }=0$.
\end{description}

\noindent \textbf{Proof:} The proof of parts (i)-(iii) are obvious.
For part (iv), we have

\[0\leq{\left(1-{\exp  \left(-ax-\frac{b}{2}x^2\right)\ }\right)}^{\alpha-1}<1\Rightarrow 0<f\left(x\right)<\ \frac{a+bx}{B\left(\alpha ,\beta \right)}{\exp  \left(-a\beta x-\frac{b\beta}{2}x^2\right)\
}.\] It can be easily shown that
\[{\mathop{\lim }_{x\rightarrow\infty} (a+bx){\exp  \left(-a\beta x-\frac{b \beta}{2}x^2\right) } }=0. \]
and the proof is completed. $\hfill\blacksquare$

\bigskip

\noindent {\bf Theorem 2.} Let $f(x)$ be the density function of the BLFR distribution. The mode of $f$ is given in the following cases:
\begin{description}
\item[i.] If $\alpha =1$ and $-a+\sqrt{\frac{b}{\beta }\ }>0$ then $f(x)$ has a unique mode in $x=\frac{1}{b}(-a+\sqrt{\frac{b}{\beta }\ })$

\item[i.] If $\alpha =1$ and  $-a+\sqrt{\frac{b}{\beta }\ }<0$ then $f(x)$ has a unique mode in $x=0$.

\item[ii.] If $\alpha >1$ then $f(x)$ has at least one mode.

\end{description}

\noindent \textbf{Proof:} The proof is obvious and is omitted.
$\hfill\blacksquare$

\bigskip

\noindent {\bf Theorem 3.} Let $h(x)$ be the hazard rate function of
the BLFR distribution. Consider the following cases:

\begin{description}
\item[i.] If $\alpha =1$ and $b>0$ then BLFR distribution has an increasing hazard rate function.

\item[ii.] If $\alpha > 1$ and $b>0$ then the hazard rate function of the BLFR distribution is an increasing.

\item[iii.] If $b = 0$ BLFR distribution has a decreasing hazard rate function for $\alpha < 1(>1)$, and $h(x)$ is constant for $\alpha =
1$.

\item[iv.] If $\alpha < 1$ and $b>0$ then $h(x)$ is a bathtub-shaped.
\end{description}

\noindent \textbf{Proof:}

\noindent i. If $\alpha =1$ then $B\left(\alpha ,\beta \right)-B_{G(x)}\left(\alpha ,\beta \right)=\frac{1}{\beta}\ {\exp  \left(-a\beta x-\frac{b\beta }{2}x^2\right)\ }$. Therefore
\[h\left(x\right)=\frac{a+bx}{\frac{1}{\beta}\ {\exp  \left(-a\beta x-\frac{b\beta }{2}x^2\right)\ }}{\exp  \left(-a\beta x-\frac{b\beta }{2}x^2\right)\
}=�\left(a+bx\right),\] which is an increasing and linear function
with respect to $x.$

\noindent ii. Consider
\[z=ax+\frac{b}{2}x^2=\frac{b}{2}{\left(x+\frac{a}{b}\right)}^2-\frac{a^2}{2b}\]
It implies that $z>0$ for $x>0$ and also, it is increasing with respect to $x$. We have $x=\frac{1}{b}\sqrt{2bz+a^2}-\frac{a}{b}$. Now, rewriting the BLFR density as function of $z$, $\xi \left(z\right)$ say, we obtain
\[\xi \left(z\right)=f\left(\sqrt{2bz+a^2}-\frac{a}{b}\right)=\frac{\sqrt{2bz+a^2}}{B\left(\alpha ,\ \beta \right)}{\left(1-{\exp  \left(-z\right)\ }\right)}^{\alpha -1}{\exp  \left(-\beta z\right).}\]
Therefore, we have
\[\frac{{\partial }^2}{\partial z^2}{\log  \xi \left(z\right)\ }=-2b^2{\left(2bz+a^2\right)}^{-2}+\left(\alpha -1\right)\frac{-{\exp  \left(-z\right)\ }}{\ {\left(1-{\exp  \left(-z\right)\ }\right)}^2}<0,\]
and we conclude that the hazard function of BLFR distribution is increasing.

\noindent iii. If $b=0$ then
\[{\log  \left(f\left(x\right)\right)\ }={\log  \left(a\right)\ }-{\log  \left(B\left(\alpha ,\ \beta \right)\right)+\left(\alpha -1\right)\ }{\log  \left(1-{\exp  \left(-ax\right)\ }\right)\ }-a\beta x,\]
and
\[\frac{{\partial }^2}{\partial x^2}{\log  \left(f\left(x\right)\right)\ }=-\frac{\left(\alpha -1\right){\exp  \left(-ax\right)\ }}{\left(1-{\exp  \left(-ax\right)\ }\right)^2}.\]
Thus we have $\frac{{\partial }^2}{\partial x^2}{\log
\left(f\left(x\right)\right)\ }>0(<0)$ where $\alpha<1(>1)$, which
implies the decreasing (increasing) hazard rate functions in this
cases.

\noindent iv. It is difficult to determine analytically the regions
corresponding to the upside-down bathtub shaped (unimodal) and
bathtub-shaped hazard rate functions for the BLFR distribution.
However, by some graphical analysis we can shows: bathtub-shaped
hazard rate function correspond to $\alpha < 1$ and $b>0$. the proof
is completed. $\hfill\blacksquare$

\section{Some extensions and Moments of the BLFR distribution}

Here, we present some representations of CDF, PDF, and the survival function of BLFR distribution. The mathematical relation given below will be useful in this section. If $\beta $ is a positive real non-integer and $\left|z\right|<1$, then
\[{\left(1-z\right)}^{\beta -1}=\sum^{\infty }_{j=0}{w_jz^j},\]
and if $\beta $ is a positive real integer, then the upper of the this summation stops at $\beta-1$, where
$$w_j=\frac{{\left(-1\right)}^j\Gamma (\beta )}{\Gamma (\beta -j)\Gamma (j+1)}.$$

\noindent 1. We can express (\ref{eq.FB}) as a mixture of distribution function of generalized LFR distributions as follows:
\[F\left(x\right)=\sum^{\infty }_{j=0}{p_j{\left(G\left(x\right)\right)}^{\alpha +j}}=\sum^{\infty }_{j=0}{p_jG_j(x)},\]
where $p_j=\frac{{\left(-1\right)}^j\Gamma (\alpha +\beta )}{\Gamma\left(\alpha \right)\Gamma \left(\beta -j\right)\Gamma\left(j+1\right)\left(\alpha +j\right)}$ and $G_{{j}}\left(x\right)={\left(G\left(x\right)\right)}^{\alpha +j}$ is distribution function of a random variable which has a generalized
LFR distribution with parameters $a$, $b$, and $\alpha +j$.

\noindent 2. We can express (\ref{eq.fBLFR}) as a mixture of density function of generalized LFR distributions as follows:
\[f\left(x\right)=\sum^{\infty }_{j=0}{p_j{\rm (}\alpha {\rm +}j{\rm )g}\left(x\right){\left(G\left(x\right)\right)}^{\alpha +j-1}}=\sum^{\infty }_{j=0}{p_j{{\rm g}}_{j}\left(x\right)},\]
where ${{\rm g}}_{j}\left(x\right)$ is density function of a random variable which has a generalized LFR distribution with
parameters $a$, $b$, and $\alpha +j$.

\noindent 3. The cdf  can be expressed in terms of the hypergeometric function and the incomplete beta function ratio (see, Cordeiro and Nadarajah, 2011) in the following way:
\[F\left(x\right)=\frac{{G\left(x\right)}^{\alpha }}{\alpha B\left(\alpha ,\beta \right)}\ _2F_1\left(\alpha ,1-\beta ;\alpha +1;G\left(x\right)\right),\]
where $_2F_1\left(a,b;c;z\right)=\sum^{\infty }_{k=0}{\left({\left(a\right)}_k{\left(b\right)}_k\right)/\left({\left(c\right)}_kk!\right)z^k}$.

\noindent 4. The $k$th moment of BLFR distribution can be expressed as a mixture of the $k$th moment of generalized LFR distributions as follows:
\begin{eqnarray}
E(X^k)&=&\int^{\infty }_0{x^kf\left(x\right)dx}=\int^{\infty }_0{x^k\sum^{\infty }_{j=0}{p_j{\rm (}\alpha {\rm +}j{\rm )g}\left(x\right){\left(G\left(x\right)\right)}^{\alpha +j-1}}dx} \nonumber\\
&=&\sum^{\infty }_{j=0}{p_j\int^{\infty }_0{x^k{g}_{j}\left(x\right)dx}}=\sum^{\infty }_{j=0}{p_jE(X^k_j)},
\end{eqnarray}
where ${g}_{j}\left(x\right)$ is density function of a random variable $X_j$ which has a generalized LFR distribution with parameters $a$, $b$, and $\alpha +j$.

\section{ Estimation and inference}

Consider $X_1,\dots X_n$ is a random sample from BLFR distribution.  The log-likelihood function for the vector of parameters $\mbox{\boldmath $\theta$}=(a,b,\alpha ,\beta )$ can be written as
\begin{eqnarray}\label{eq.like}
\ell \left(\mbox{\boldmath $\theta$}\right)&=&\sum^n_{i=1}{{\log \left(a+bx_i\right) }}-n{\log  \left(\Gamma \left(\alpha \right)\right) }-n{\log  \left(\Gamma \left(\beta \right)\right)}\nonumber \\
&&+n{\log \left(\Gamma \left(\alpha +\beta \right)\right)}+\left(\alpha-1\right)\sum^n_{i=1}{{\log \left(1-{\exp  \left(t_i\right)\ }\right)}}+\beta\sum^n_{i=1}{ t_i},
\end{eqnarray}
where $t_i=-a x_i-\frac{b}{2}x^2_i$. The log-likelihood can be maximized either directly or by solving the nonlinear likelihood equations obtained by differentiating (\ref{eq.like}). The components of the score vector $U\left(\mbox{\boldmath $\theta$}\right)$\textit{ }are given by
\begin{eqnarray*}
&&U_a\left(\mbox{\boldmath $\theta$}\right)=\frac{\partial }{\partial a}\ell \left(\mbox{\boldmath $\theta$}\right)=\sum^n_{i=1}{\frac{1}{a+bx_i}}+\left(\alpha-1\right)\sum^n_{i=1}{\frac{x_i{\exp \left(t_i\right) }}{1-{\exp \left(t_i\right) }}}-\beta\sum^n_{i=1}{x_i},\\
&&
U_b\left(\mbox{\boldmath $\theta$}\right)=\frac{\partial }{\partial b}\ell \left(\mbox{\boldmath $\theta$}\right)=\sum^n_{i=1}{\frac{x_i}{a+bx_i}}+\frac{\left(\alpha-1\right)}{2}\sum^n_{i=1}{\frac{x^2_i{\exp \left(t_i\right)\ }}{1-{\exp  \left(t_i\right)\ }}}-\frac{\beta}{2}\sum^n_{i=1}{x^2_i},\\
&&
U_{\alpha}\left(\mbox{\boldmath $\theta$}\right)=\frac{\partial }{\partial\alpha }\ell \left(\mbox{\boldmath $\theta$}\right)=-n\psi \left(\alpha\right)+n\psi \left(\alpha+\beta\right)+\sum^n_{i=1}{{\log  \left(1-{\exp  \left(t_i\right)}\right)\ }},\\
&& U_{\beta }\left(\mbox{\boldmath $\theta$}\right)=\frac{\partial }{\partial\beta }\ell \left(\mbox{\boldmath $\theta$}\right)=-n\psi
\left(\beta\right)+n\psi \left(\alpha+\beta\right)
+\sum^n_{i=1}{t_i}.
\end{eqnarray*}
where $\psi \left(.\right)$ is the digamma function.

For interval estimation and hypothesis tests on the model parameters, we require the observed information matrix. The $4\times 4$ unit observed information matrix $J=J(\mbox{\boldmath $\theta$})$ is obtained as
\[J=\left[ \begin{array}{cccc}
J_{aa} & J_{ab} & J_{a\alpha } & J_{a\beta } \\
J_{ba} & J_{bb} & J_{b\alpha } & J_{b\beta } \\
J_{\alpha a} & J_{\alpha b } & J_{\alpha \alpha } & J_{\alpha \beta } \\
J_{\beta a } & J_{\beta b } & J_{ \beta \alpha} & J_{\beta \beta } \end{array}
\right].\]
where the expressions for the elements of $J$ are
\begin{align*}
&J_{aa}=\frac{{\partial }^2}{\partial a\partial a}\ell \left(\mbox{\boldmath $\theta$}\right)=-\sum^n_{i=1}{\frac{1}{{\left(a+bx_i\right)}^2}}+\left(\alpha-1\right)\sum^n_{i=1}{\frac{x^2_i{\exp  \left(t_i\right)\ }}{1-{\exp  \left(t_i\right)\ }}}-\left(\alpha-1\right)\sum^n_{i=1}{\frac{x^2_i{\exp  \left(2t_i\right)\ }}{{\left(1-{\exp  \left(t_i\right)\ }\right)}^2\ }}\\
&J_{ab}=J_{ba}=\frac{{\partial }^2}{\partial b\partial a}\ell \left({\mbox{\boldmath $\theta$} }\right)=-\sum^n_{i=1}{\frac{x_i}{{\left(a+bx_i\right)}^2}}+\frac{\left(\alpha-1\right)}{2}\sum^n_{i=1}{\frac{x^3_i{\exp  \left(t_i\right)\ }}{1-{\exp  \left(t_i\right)\ }}}-\frac{\left(\alpha-1\right)}{2}\sum^n_{i=1}{\frac{x^3_i{\exp  \left(2t_i\right)\ }}{{\left(1-{\exp  \left(t_i\right)\ }\right)}^2\ }}\\
&J_{a\alpha }=J_{\alpha a}=\frac{{\partial }^2}{\partial\alpha \partial a}\ell \left({\mbox{\boldmath $\theta$} }\right)=\sum^n_{i=1}{\frac{x_i{\exp  \left(t_i\right)\ }}{1-{\exp  \left(t_i\right)\ }}}
\end{align*}
\begin{align*}
&J_{a\beta }=J_{\beta a}=\frac{{\partial }^2}{\partial\beta \partial a}\ell \left({\mbox{\boldmath $\theta$} }\right)=\sum^n_{i=1}{x_i}\\
&J_{bb}=\frac{{\partial }^2}{\partial b\partial b}\ell \left({\mbox{\boldmath $\theta$} }\right)=-\sum^n_{i=1}{\frac{x^2_i}{{\left(a+bx_i\right)}^2}}-\frac{\left(\alpha-1\right)}{4}\sum^n_{i=1}{\frac{x^4_i{\exp  \left(t_i\right)\ }}{1-{\exp  \left(t_i\right)\ }\ }}-\frac{\left(\alpha-1\right)}{4}\sum^n_{i=1}{\frac{x^4_i{\exp  \left(2t_i\right)\ }}{{\left(1-{\exp  \left(t_i\right)\ }\right)}^2\ }}\\
&J_{b\alpha }=J_{\alpha b}=\frac{{\partial }^2}{\partial\alpha \partial b}\ell \left({\mbox{\boldmath $\theta$} }\right)=\frac{1}{2}\sum^n_{i=1}{\frac{x^2_i{\exp  \left(t_i\right)\ }}{1-{\exp  \left(t_i\right)\ }}}\\
&J_{b\beta }=J_{\beta b}=\frac{{\partial }^2}{\partial \beta \partial b}\ell \left({\mbox{\boldmath $\theta$} }\right)=
-\frac{1}{2}\sum^n_{i=1}{x_i}\\
&J_{\alpha\alpha }=\frac{{\partial }^2}{\partial \alpha^2}\ell \left({\mbox{\boldmath $\theta$} }\right)=n{\psi}'\left(\alpha+\beta\right)-n{\psi}'\left(\alpha\right)\\
&J_{\alpha \beta }=J_{\beta \alpha }=\frac{{\partial }^2}{\partial\beta \partial\alpha }\ell \left({\mbox{\boldmath $\theta$} }\right)=n{\psi}'\left(\alpha+\beta\right)\\
&J_{\beta \beta }=\frac{{\partial }^2}{\partial\beta \partial\beta }\ell \left({\mbox{\boldmath $\theta$} }\right)=n{\psi }'\left(\alpha+\beta\right)-n{\psi }'\left(\beta\right)
\end{align*}
where ${\psi }'\left(.\right)$ is the trigamma function.

Under conditions that are fulfilled for parameters in the interior of the parameter space but not on the boundary, asymptotically
\[\sqrt{n}\left(\widehat{\mbox{\boldmath $\theta$}}-\mbox{\boldmath $\theta$}\right)\sim N_4\left(0,\ I{\left(\mbox{\boldmath $\theta$}\right)}^{-1}\right),\]
where $I\left(\mbox{\boldmath $\theta$}\right)$\textit{ }is the expected information matrix. This asymptotic behavior is valid if $I\left(\mbox{\boldmath $\theta$}\right)$ is replaced by $J(\widehat{\mbox{\boldmath $\theta$}})$, i.e., the observed information matrix evaluated at $\widehat{\mbox{\boldmath $\theta$}}$.

 For  constructing  tests of hypothesis and confidence region we can use from this result. An asymptotic confidence interval with confidence level $1-\gamma $ for each parameter $\theta_i$, $i=1,2,3,4$, is given by
 $$\left( \hat{\theta}_i-z_{\gamma/2}\sqrt{J^{\theta_i}} \ , \ \hat{\theta}_i+z_{\gamma/2}\sqrt{J^{\theta_i}}\right),$$
where $J^{\theta_i}$ is the $i$th diagonal element of $J(\widehat{\mbox{\boldmath $\theta$}})$ and $z_{\gamma/2}$ is the upper $\gamma/2$ point of standard normal distribution.

\section{Application of BLFR to real data set}

In this section, we provide a data analysis to see how the new model works in practice. This data set is given by Aarset (1987) and
consists of times to first failure of fifty devices. The data is given by

0.1, 0.2, 1, 1, 1, 1, 1, 2, 3, 6, 7, 11, 12, 18, 18, 18, 18, 18, 21, 32, 36, 40, 45, 46, 47, 50,

55, 60, 63, 63, 67, 67, 67, 67, 72, 75, 79, 82, 82, 83, 84, 84, 84, 85, 85, 85, 85, 85, 86, 86.

In this section we fit BLFR, GLFR, LFR, GR, GE, Rayleigh and
exponential models to the above data set. We use the maximum
likelihood method to estimate the model parameters and
 calculate the standard errors of the MLE's, respectively. The MLEs of the
parameters (with std.), the maximized log-likelihood, the
Kolmogorov-Smirnov statistic with its respective \textit{p}-value,
the AIC (Akaike Information Criterion), AICC and BIC (Bayesian
Information Criterion) for the BLFR, GLFR, LFR, GR, GE, Rayleigh and
exponential models are given in Table \ref{table1}.

We can perform formal goodness-of-fit tests in order to verify which
distribution fits better to the first data. We apply the
Anderson-Darling (AD) and Cramr�von Mises (CM) tests. In general,
the smaller the values of AD and CM, the better the fit to the data.
For this data set, the values of AD and CM statistics for fitted
distributions are given in in Table \ref{table1}.

\begin{figure}[]
\centering
\includegraphics[scale=0.45]{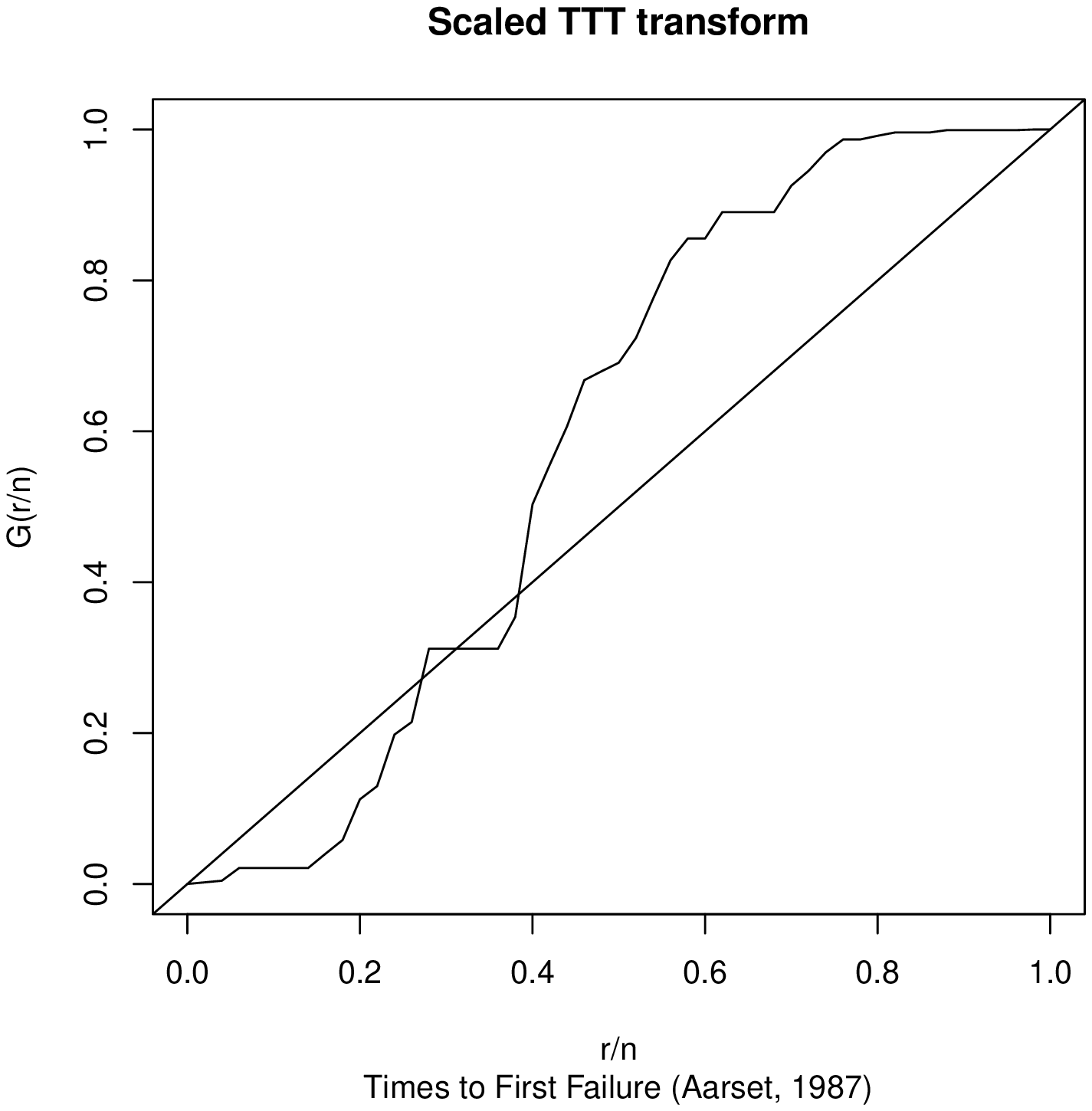}
\includegraphics[scale=0.46]{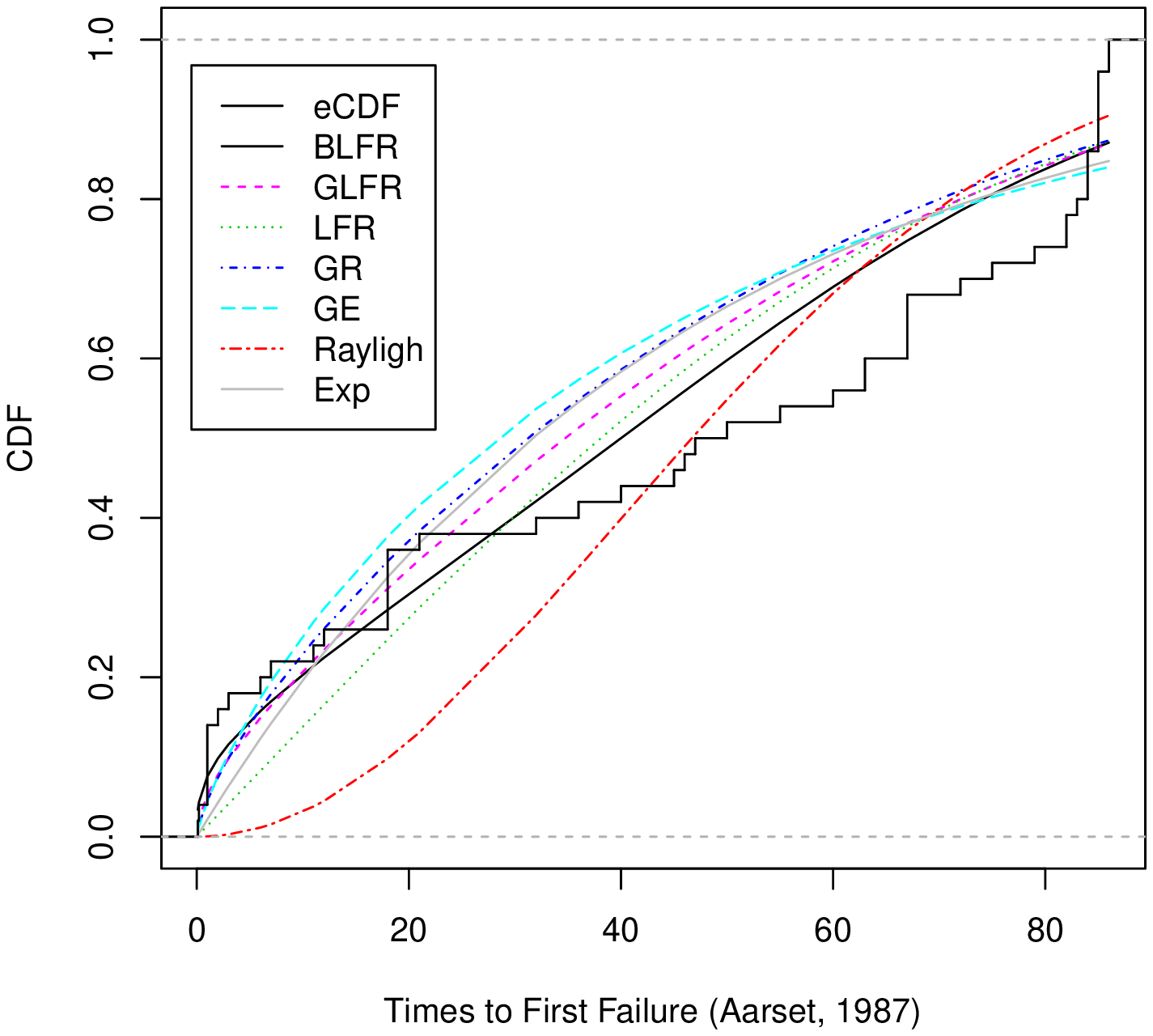}
\caption[]{Plots of the empirical scaled TTT transform (left), and the empirical distribution versus the fitted cumulative distribution functions (right). } \label{fig.ex1}
\end{figure}

The empirical scaled TTT transform (Aarset, 1987) can be used to identify the shape of the hazard function. The scaled TTT transform
is convex (concave) if the hazard rate is decreasing (increasing), and for bathtub (unimodal) hazard rates, the scaled TTT transform is
first convex (concave) and then concave (convex). The TTT plot for this data in Fig. \ref{fig.ex1} shows a bathtub-shaped hazard rate function and,
therefore, indicates the appropriateness of the BLFR distribution to fit this data. The empirical distribution versus the fitted cumulative distribution functions of BLFR, GLFR, LFR, GR, GE, Rayleigh and exponential distributions are displayed in Fig. \ref{fig.ex1}.

The results for this data set show that the BLFR distribution yields the best fit among the GLFR, LFR, GR, GE, Rayleigh and exponential
distributions.  For this data, the K-S test statistic takes the smallest value with the largest value of its respective
\textit{p}-value for BLFR distribution. Also this conclusion is confirmed from the values of the AIC, AICC and BIC for the fitted
models given in Table \ref{table1} and the plots of the densities and cumulative distribution functions in Fig. \ref{fig.ex1}.

Using the likelihood ratio (LR) test, we test the null hypothesis $H_0$: GLFR versus the alternative hypothesis H1: BLFR, or
equivalently, $H_0$: $b=0$ versus $H_1$: $b\neq 0$. The value of the LR test statistic and the corresponding \textit{p}-value are 3.4 and
0.019, respectively. Therefore, the null hypothesis (GLFR model) is rejected in favor of the alternative hypothesis (BLFR model) for a
significance level $>$ 0.019. For test the null hypothesis $H_0$: LFR versus the alternative hypothesis $H_1$: BLFR, or equivalently, $H_0$:
$(a,b)=(1,1)$ versus $H_1$: $(a,b)\neq(1,1)$, the value of the LR test statistic is 15.3 (\textit{p}-value = 0.00047), which includes that
the null hypothesis (LFR model) is rejected in favor of the alternative hypothesis (BLFR model) for any significance level. We
also test the null hypothesis $H_0$: GR versus the alternative hypothesis $H_1$: BLFR, or equivalently, $H_0$: $(\alpha,b)=(1,1)$ versus
$H_1$: $(\alpha,b)\neq(1,1)$. The value of the LR test statistic is 8.3 (\textit{p}-value = 0.0158), which includes that the null hypothesis
(GR model) is rejected in favor of the alternative hypothesis (BLFR model) for a significance level $> 0.0158$. For test the null
hypothesis $H_0$: GE versus the alternative hypothesis $H_1$: BLFR, or equivalently, $H_0$: $(\beta,b)=(1,1)$ versus $H_1$: $(\beta,b)\neq(1,1)$,
the value of the LR test statistic is 19.2 (\textit{p}-value = 6e-05), which includes that the null hypothesis (GR model) is
rejected in favor of the alternative hypothesis (BLFR model) for any significance level.

\section{Simulations}
This section provides the results of simulation study. simulations
have been performed in order to investigate the proposed estimator
of $\alpha ,$ $\beta $,$a$ and $b$ of the proposed MLE method. We
generated 10000 samples of size $n=30,\ 50,\ 100$ and $200$ from the
BLFR distribution for each one of the six set of values of
$\left(\alpha ,\beta ,a,b \right).$ We assess the accuracy of the
approximation of the standard error of the MLEs determined though
the Fisher information matrix. The approximate values of
$se(\hat{\alpha })$, $se(\hat{\beta })$, $se(\hat{a})$ and
$se(\hat{b})$ are computed. The results for the BLFR distribution is
shown in Table \ref{table.2}, which indicate the following results: (i)
convergence has been achieved in all cases and this emphasizes the
numerical stability of the MLE method. (ii) The differences between
the average estimates and the true values are almost small. (iii)
These results suggest that the MLE estimates have performed
consistently. (iv) The standard errors of the MLEs decrease when the
sample size increases.

\section{ Conclusion}
 We define a new model, called the BLFR distributions, which generalizes the LFR and GLFR distributions. The BLFR distributions contain the GLFR, LFR,GR, GE, Rayleigh and exponential distributions as special cases. The BLFR distribution present hazard functions with a very flexible behavior. We obtain closed form
expressions for the moments. Maximum likelihood estimation is discussed. Finally, we fitted BLFR distribution to a real data set to show the potential of the new proposed class.

\begin{table}[h]
\caption{MLEs (STDs) of the fitted distribution, K-S,
\textit{p}-values, $-2\log(L)$, AIC, AICC, BIC, AD and CM
corresponds to times to first failure.}\label{table1}
\begin{center}
\begin{tabular}{|c|ccccccc|} \hline
 &  \multicolumn{7}{|c|}{Distribution}  \\ \hline
            & \textbf{BLFR} & \textbf{GLFR}   & \textbf{LFR} & \textbf{GR} & \textbf{GE} & \textbf{Rayleigh} & \textbf{Exp.} \\ \hline
$\hat{a}$   & 0.3347   & 0.5327   & ---          & 0.3520      & 0.7798      & ---               & --- \\
      (s.e.)      & (0.1432) & (0.1145) & ---          & (0.0559)    & (0.1351)    & ---               & --- \\ \hline

$\hat{b}$    & 0.1243   & ---      & ---          & ---         & ---         & ---               & --- \\
       (s.e.)     & (0.0722) & ---      & ---          & ---         & ---         & ---               & --- \\ \hline

$\hat{\alpha}$ & 0.0172   & 0.0038   & 0.0136    & ---    & 0.0187    & --- & 0.02189 \\
         (s.e.)      & (0.0354) & (0.0030) & (0.0038)  & ---    & (0.00363) & --- & (0.00309) \\ \hline

$\hat{\beta}$  & 0.0348   & 0.00031   & 0.00024  & 0.00031   & --- &  0.00064  & --- \\
        (s.e.)       & (0.0025) & (0.00008) & (0.0001) & (0.00008) & --- & (0.00009) & --- \\ \hline \hline

-2log L  & 460.8  & 466.3  & 476.1  & 469.1  & 480.0  & 528.1  & 482.2  \\
AIC  & 468.8  & 472.3  & 480.1  & 473.1  & 484.0  & 530.1  & 484.2  \\
AICC  & 469.6  & 472.8  & 480.4  & 473.4  & 484.2  & 530.2  & 484.3  \\
BIC  & 476.4  & 478.0  & 484.0  & 477.0  & 487.8  & 532.0  & 486.1  \\
K-S & 0.1554 & 0.1830 & 0.1768 & 0.2009 & 0.2042 & 0.2621 & 0.1911 \\
P-value & 0.1786 & 0.0703 & 0.0877 & 0.0353 & 0.0309 & 0.0021 & 0.0519 \\
AD & 1.749 & 2.4890 & 4.0346 & 3.0923 & 3.2530 & 13.3205 & 3.6505 \\
CM & 0.3574 & 0.4959 & 0.5443 & 0.6111 & 0.6472 & 0.8728 & 0.6006 \\ \hline
\end{tabular}
\end{center}
\end{table}

\begin{table}[]
\begin{center}
\caption{The averages of the 10000 MLE's and mean of the simulated standard errors for BLFR distribution. }\label{table.2}
{\small
\begin{tabular}{|c|c|cccc|cccc|} \hline
 &  &  \multicolumn{4}{|c|}{AE}  &  \multicolumn{4}{|c|}{SD}   \\ \hline
$n$ & $(\alpha ,\beta ,a,b)$ & $\widehat{\alpha }$ & $\widehat{\beta }$ & $\hat{a}$ & $\hat{b}$ & $sd(\widehat{\alpha })$ & $sd(\widehat{\beta })$ & $sd(\hat{a})$ & $sd(\hat{b})$  \\ \hline
30 & (.5,.5,1,1) & 0.534 & 0.537 & 2.453 & 2.363 & 0.449 & 0.606 & 4.346 & 3.454   \\
   & (.5,.5,1,2) & 0.528 & 0.671 & 1.896 & 4.485 & 0.290 & 0.875 & 3.191 & 5.528    \\
   & (.5,.5,3,1) & .4915 & .7123 & 3.321 & 3.082 & 0.145 & 0.571 & 5.336 & 5.236   \\
   & (1,2,1,3)   & 1.158 & 2.058 & 1.977 & 5.266 & 0.534 & 1.415 & 3.576 & 6.547   \\
   & (3,2,1,1)   & 4.995 & 4.043 & 1.275 & 2.623 & 8.496 & 2.761 & 1.754 & 7.102   \\
   & (3,3,3,3)   & 3.251 & 4.035 & 3.249 & 3.568 & 1.627 & 2.054 & 3.520 & 3.978  \\
\hline
50 & (.5,.5,1,1) & 0.497 & 0.714 & 1.623 & 1.705 & 0.151 & 0.843 & 2.505 & 2.153   \\
   & (.5,.5,1,2) & 0.504 & 0.709 & 1.798 & 3.720 & 0.170 & 0.934 & 2.233 & 4.437    \\
   & (.5,.5,3,1) & .491  & 0.730 & 3.579 & 1.937 & .1165 & .5404 & 6.525 & 2.658    \\
   & (1,2,1,3)   & 1.079 & 2.053 & 1.658 & 5.232 & .3852 & 1.392 & 2.272 & 6.470   \\
   & (3,2,1,1)   & 4.308 & 4.237 & 1.222 & 2.635 & 6.246 & 2.982 & 1.654 & 7.820   \\
   & (3,3,3,3)   & 3.086 & 3.874 & 3.131 & 3.859 & 1.307 & 1.923 & 3.073 & 4.716    \\
   \hline
100 & (.5,.5,1,1) & .4916 & .7806 & 1.683 & 1.293 & .1087 & .9284 & 2.698 & 1.179   \\
    & (.5,.5,1,2) & .4892 & .7829 & 3.597 & 3.547 & 3.277 & .2104 & 5.221 & 3.875  \\
    & (.5,.5,3,1) & .4923 & .7513 & 3.741 & 1.275 & .0815 & .5388 & 5.177 & 1.556  \\
    & (1,2,1,3)   & 1.023 & 2.169 & 1.396 & 5.047 & .2841 & 1.443 & 1.470 & 6.017  \\
    & (3,2,1,1)   & 3.385 & 4.122 & 1.079 & 2.169 & 2.669 & 2.977 & 1.339 & 4.954  \\
    & (3,3,3,3)   & 2.998 & 3.779 & 3.091 & 3.612 & .8535 & 1.973 & 2.374 & 4.357  \\
\hline
200 & (.5,.5,1,1) & .4897 & .7999 & 1.465 & 1.121 & .0739 & .9091 & 2.022 & .8251  \\
    & (.5,.5,1,2) & .4881 & .7773 & 1.642 & 2.495 & .0845 & .9556 & 2.227 & 1.902  \\
    & (.5,.5,3,1) & .4936 & .7369 & 3.762 & 1.036 & .0567 & .4984 & 4.427 & 1.038  \\
    & (1,2,1,3) & 1.005 & 2.062 & 1.264 & 4.522 & .1967 & 1.122 & .8104 & 4.719 \\
    & (3,2,1,1) & 3.070 & 4.021 & .9440 & 1.987 & 1.311 & 2.944 & .9673 & 3.587  \\
    & (3,3,3,3) & 2.999 & 3.766 & 3.053 & 3.626 & .6332 & 2.025 & 1.940 & 3.962  \\ \hline
\end{tabular}
}
\end{center}
\end{table}


\begin{thebibliography}{99}


\bibitem{} Aarset, M. V. (1987). How to identify bathtub hazard rate, \textit{IEEE Transactions on Reliability}, 36(1), 106--108.

\bibitem{} Akinsete, A., Famoye, F., Lee, C. (2008). The beta-Pareto distribution, \textit{Statistics}, 42(6), 547--563.  

\bibitem{} Akinsete, A., Lowe C. (2009). Beta-Rayleigh distribution in reliability measure. Section on Physical and Engineering Sciences, \textit{Proceedings of the American Statistical Association}, (1), 3103--3107.


\bibitem{} Barreto-Souza, W., Santos A. H. S., Cordeiro, G. M. (2010). The beta generalized exponential distribution, {\it Journal of Statistical Computation and Simulation}, 80, 159--172.

\bibitem{} Barreto-Souza, W., Cordeiro, G.,  Simas, A. (2011). Some Results for Beta Frechet Distribution, {\it Communication in Statistics-Theory and Methods}, 40(5), 798-811.

\bibitem{}  Cordeiro, G. M., Simas, A. B., Stosic, B. (2008). Explicit expressions for moments of the beta Weibull distribution, Preprint:/arXiv:0809.1860v1S.

\bibitem{} Cordeiroa, G. M., Silva, G. O., Ortega, E. E. M.   A. D. C., Cintra, L. C., Rago, L. C. (2011). The beta-Weibull geometric distribution, {\it Statistics}, DOI: 10.1080/02331888.2011.577897.

\bibitem{} Cordeiroa, G. M., Gomez, A. E., de Silva, C. Q., Ortega, E. E. M. (2011). The beta exponentiated Weibull distribution, {\it Journal of Statistical Computation and Simulation}, DOI: 10.1080/00949655.2011.615838.

\bibitem{}  Cordeiro, G. M.,   Nadarajah, S. (2011). Closed form expressions for moments of a class of Beta generalized distributions, {\it Brazilian Journal of Probability and Statistics}, 25, 14--33.

\bibitem{} Cordeiroa, G. M., Castellaresb, F., Montenegrob, L. C., de Castro, M. (2012). The beta generalized gamma distribution, {\it Statistics}, DOI: 10.1080/02331888.2012.658397.

\bibitem{} Cordeiroa, G. M., Nascimento,  A. D. C., Cintra, L. C., Rago, L. C. (2012). Beta generalized normal distribution with an application for SAR image processing, {\it Statistics}, DOI: 10.1080/02331888.2012.748776.

\bibitem{}  Eugene, N., Lee, C., Famoye, F. (2002). Beta-normal distribution and its applications, {\it Communication in Statistics-Theory and Methods}, 31, 497--512.

\bibitem{}  Famoye, F., Lee, C., Olumolade, O. (2005). The beta-Weibull distribution, {\it Journal of Statistical Theory and Applications}, 4 (2), 121--136.


\bibitem{}  Gupta, R. D., Kundu, D. (1999). Generalized exponential distribution, {\it Australian and New Zealand Journal of Statistics}, 41 (2), 173--188.

\bibitem{} Khan, M. S. (2010). The beta inverse Weibull distribution, {\it International Transactions in Mathematical Sciences and Computer}, 3, 113-119.

\bibitem{} Kundu, D., Raqab, M. (2005). Generalized Rayleigh distribution: different methods of estimations, {\it Computational Statistics and Data Analysis}, 49, 187--200.

\bibitem{} Lee, C., Famoye, F., Olumolade, O. (2007). Beta-Weibull distribution: some properties and applications to censored data, {\it Journal of Modern Applied Statistical Methods}, 6, 173--86.

\bibitem{} Mahmoudi, E. (2011). The beta generalized Pareto distribution with application to lifetime data, {\it Mathematics and Computers in Simulation}, 81(11),  2414--2430.

\bibitem{} Nadarajah, S., Gupta, A. K. (2004). The beta Fr$\acute{{\rm e}}$chet distribution, {\it Far East Journal of Theoretical Statistics}, 14, 15--24.

\bibitem{} Nadarajah, S., Kotz, S. (2004). The beta Gumbel distribution, {\it Mathematical Problems in Engineering}, 10, 323--332.

\bibitem{}  Nadarajah, S., Kotz, S. (2006). The beta exponential distribution, {\it Reliability Engineering and System Safety}, 91, 689--697.

\bibitem{}  Sarhan, M., Kundu, D. (2009). Generalized Linear Failure Rate Distribution, {\it Communications in Statistics-Theory and Methods}, 38(5), 642--660.

\bibitem{}  Sen, A., Bhattacharya, G. K. (1995). Inference procedure for the linear failure rate model, {\it Journal of Statistical Planning and Inference}, 46, 59--76.

\bibitem{}  Silva, G. O., Ortega, E. M. M., Cordeiro, G. M. (2010). The beta modified Weibull distribution, {\it Lifetime Data Analysis}, 16, 409--430.

\bibitem{} Singla, N., Jain, K., Kumar Sharma, S. (2012). The Beta Generalized Weibull distribution: Properties and applications, {\it Reliability Engineering} \& {\it System Safety}, 102, 5--15.

  \bibitem{} Surles, J. G., Padgett, W. J. (2005). Some properties of a scaled Burr type X distribution, {\it Journal of Statistical Planning and Inference}, 128, 271--280.

\end{thebibliography}
\end{document}